\documentclass[a4paper, aps, prr, amsfonts, amssymb, amsmath, reprint, twoside, superscriptaddress,longbibliography,nofootinbib]{revtex4-2}
\usepackage[english]{babel}
\usepackage[utf8]{inputenc}
\usepackage{comment}
\usepackage{siunitx}
\AtBeginDocument{\RenewCommandCopy\qty\SI}
\usepackage{amsthm}
\usepackage{mathtools}
\usepackage{physics}
\usepackage{xcolor}
\usepackage{graphicx}
\usepackage[left=23mm,right=13mm,top=35mm,columnsep=15pt]{geometry} 
\usepackage{adjustbox}
\usepackage{placeins}
\usepackage[T1]{fontenc}
\usepackage{csquotes}
\usepackage{tabularray}
\usepackage[pdftex, pdftitle={Article}, pdfauthor={Author}]{hyperref}
\begin{document}
\title{A framework for the evaluation of transport-induced excitations in shuttling-based ion-trap quantum processors}

\author{Rodrigo Munoz}
\thanks{These authors contributed equally to this work.}
\email[]{munoz@iqo.uni-hannover.de}
\affiliation{Institute for Quantum Optics, Leibniz University Hannover, Welfengarten 1, Hannover, 30167, Germany}

\author{Phil Nuschke}
\thanks{These authors contributed equally to this work.}
\email[]{nuschke@iqo.uni-hannover.de}
\affiliation{Institute for Quantum Optics, Leibniz University Hannover, Welfengarten 1, Hannover, 30167, Germany}

\author{Teresa Meiners}
\affiliation{Institute for Quantum Optics, Leibniz University Hannover, Welfengarten 1, Hannover, 30167, Germany}

\author{Brigitte Kaune}
\affiliation{Institute for Quantum Optics, Leibniz University Hannover, Welfengarten 1, Hannover, 30167, Germany}

\author{Christian Ospelkaus}
\affiliation{Institute for Quantum Optics, Leibniz University Hannover, Welfengarten 1, Hannover, 30167, Germany}
\affiliation{Physikalisch-Technische Bundesanstalt, Bundesallee 100, 38116 Braunschweig, Germany}

\date{\today}

\begin{abstract}
We develop a theoretical and numerical framework to analyze the effect of transport on the motional states of ions in a trapped-ion quantum processor. We decompose the shuttling protocol into primitive operations and characterize these in terms of their heating performance. Instead of having to simulate the whole transport protocol for each complete ion trajectory, the method allows us to determine the heating properties of each primitive operation separately and obtain the global result through an algebraic expression. We demonstrate our method by applying it to an 8-qubit quantum processor design based on linear transport and swap operations for all-to-all connectivity. We show how to incorporate the price of motional operations at the level of the compiler as a cost function.

\end{abstract}


\maketitle

\section{Introduction} \label{sec:outline}

Trapped ions are an experimentally advanced platform to implement fault-tolerant quantum computers~\cite{ciracQuantumComputationsCold1995,bermudezAssessingProgressTrappedIon2017}. They combine long coherence times, high-fidelity gate operations and a clear path to scalability~\cite{kielpinskiArchitectureLargescaleIontrap2002}. As system sizes grow, the total cost of operations increases, in particular regarding ion transport in the Quantum Charge-Coupled Device architecture~\cite{winelandExperimentalIssuesCoherent1998a,kielpinskiArchitectureLargescaleIontrap2002}. In this approach, universal operations are achieved by physically shuttling ions through trap arrays between specialized zones for e.g. storage, quantum gates, cooling or detection. In particular, this platform enables the implementation of an all-to-all qubit connectivity. Provided that the appropriate error correction code is selected, this can significantly reduce overhead and enable reliable error correction. Even with a small code distance, the number of physical qubits required for the implementation of a logical qubit is significantly reduced~\cite{linkeExperimentalComparisonTwo2017,cohenLowoverheadFaulttolerantQuantum2022}.

While it is common to think of the cost of computation in terms of the number of quantum gates involved~\cite{durandauAutomatedGenerationShuttling2023} or in terms of the time spent on the computation, here we analyze the motional excitation of the qubits in particular. Two-qubit gate operation imposes limits on the amount of permissible motional excitation for the participating qubits; specific requirements depend on the type of gate implementation~\cite{ciracQuantumComputationsCold1995,sorensenQuantumComputationIons1999,leibfriedExperimentalDemonstrationRobust2003}. Beyond that level, sympathetic cooling needs to be employed to re-initialize the motional state of the ions close to the ground state. Previous demonstrations of trapped-ion quantum computers used a considerable fraction of the runtime for sympathetic cooling~\cite{pinoDemonstrationTrappedionQuantum2021}. Taking motional excitations into account at the level of chip design and compilation of algorithms can therefore significantly reduce the cost of carrying out a quantum algorithm and/or improve the performance of the device. This allows for deeper insights than just optimizing the topology for a small number of transport operations, as the cost of individual transport primitives can be quite different \cite{wanIonTransportReordering2020}. The impact of transport on motional excitation has been analyzed in \cite{husimiMiscellaneaElementaryQuantum1953,huculTransportAtomicIons2008,nuschkeAnalysisMotionalHeating2025a} for single ions and in \cite{kaufmannFastIonSwapping2017,tobalinaShortcutsAdiabaticRotation2021} for ion pairs. What is still missing, however, is the integration of initial conditions in the estimation of the total phonon number after transport. In addition, it should be possible to decompose an ion rearrangement procedure at the compiler level into individual, more manageable sections for simulation. Moreover, previous studies have not taken into account the influence of the hardware setup on transport performance.

Here, we develop a method to study the cost of cascaded transport operations, by first extracting the resulting motional excitation of the individual operations starting at rest. We then examine and model the effects of initial conditions on these operations. This allows us to derive an algebraic recurrence relation for computing the phonon number of cascaded operations of arbitrary length. This represents a major reduction in computational effort, particularly for long cascaded transport sequences composed of a small number of recurring primitive operations. Such sequences are typical of QCCD architectures, and as these scale in size and complexity, the benefit becomes increasingly significant. In this way, we address the challenges outlined in the previous paragraph. 

To give a specific example, we present and study an 8-qubit processor implemented in a linear trap. We first identify the design and hardware specifications that contribute to limitations in the transport speed. Using these specifications, we find the set of trapping conditions that aim to maximize performance. After calculating the control voltages, using numerical simulations, we analyze the influence of voltage interpolation on the motional excitation.
~We feed the previously mentioned model to obtain better approximations of the total phonon number.

Together, this enables the implementation of a compiler capable of optimizing both for low heating and for transport time, providing a benchmark for our specific chip design.

In Sec.~\ref{sec:Chip Design} and Sec.~\ref{sec: trap para}, we present the trap array and discuss the critical design considerations. Furthermore, we prepare the trap to be exported into an operation compiler by discretizing the layout into a set of space intervals that the shuttling primitives need to connect. Subsequently, in Sec.~\ref{sec:single ion}, we present the methodology used to approximate transport-induced excitations and apply it to the case of single-ion transport. In Sec.~\ref{sec:merging and splitting} and Sec.~\ref{sec:swap}, we present the frameworks used to calculate the  control voltages, necessary to execute two-ion shuttling operations and extracting their contributions to heating. Finally, in Sec.~\ref{sec:Operation compiler}, and after discussing the inclusion of initial conditions in the calculation of the average phonon numbers, we present an example where the advantages of including motional heating as an operation cost is demonstrated. Moreover, an instance that could serve as benchmark for the performance of future designs is presented as well.

\begin{figure*}[t]
  \centering
  \includegraphics[width=1\textwidth]{trap_layout_NEW.png}
  \caption{\textit{a}) Electrode layout featuring a central gate zone with 14 control electrodes highlighted in red, an asymmetric RF electrode and two storage register to each side where the electrode number indicates the multiplexing of the control signals.
  Additionally, the layout includes an illustrative example of the microwave electrodes in orange. 
  The dots along the x-direction show the positions at which the ions may stay in rest. An example where all ions are stored in the left storage is shown, overlapped with the selected resting positions. \textit{b}) Register transport. Due to the multiplexing scheme, ions in the registers can only be moved collectively. \textit{c}) Merging and splitting process. Once in the gate zone, the ions can be manipulated individually. Merging means to transport two ions to the two most central positions, at which they form a two-ion crystal (shown in \textit{d})). The reversed process is called splitting. \textit{e}) Swapping: Exchanging the positions of the ions in the two-ion crystal.}
  \label{fig: layout}
\end{figure*}

\section{The chip} \label{sec:Chip Design}

The chip layout discussed in this paper has been developed within a series of chip designs to demonstrate the potential of near-field microwave quantum control (NFQC) for scalable quantum computing with trapped ions. NFQC uses chip-integrated microwave conductors~\cite{ospelkausTrappedIonQuantumLogic2008a,ospelkausMicrowaveQuantumLogic2011c} to carry out the internal-motional quantum transitions that are the essence of multi-qubit gates in the trapped-ion quantum computer~\cite{ciracQuantumComputationsCold1995}. After the initial demonstration, we focused on developing microwave chip designs that facilitate the implementation of quantum gates based on a simple and well-understood conductor structure~\cite{carsjensSurfaceelectrodePaulTrap2014,wahnschaffeSingleionMicrowaveNearfield2017}, allowing the demonstration of two-qubit gates~\cite{hahnIntegrated9BeMultiqubit2019} with high fidelity~\cite{zarantonelloRobustResourceEfficientMicrowave2019a} based on a comprehensive framework for robust amplitude-shaped pulses~\cite{duweNumericalOptimizationAmplitudemodulated2022}. Most recently, we showed that, when combined with NFQC individual-ion addressing methods based on micromotion sidebands, universal two-qubit algorithms can be implemented~\cite{pulido-mateoArbitraryQuantumCircuits2024}. Recent work at NIST~\cite{srinivasHighfidelityLaserfreeUniversal2021} has focused on developing novel NFQC gate schemes; Oxford University~\cite{hartyHighFidelityMicrowaveDrivenQuantum2013a} and Oxford Ionics / IonQ~\cite{hughesTrappedionTwoqubitGates2025a} have demonstrated low gate errors and have clearly highlighted the key advantages of NFQC: to deliver the highest fidelity quantum operations across all platforms based on fully chip-integrated control elements and with significantly reduced sensitivity to motional-state initialization.

The processor layout discussed in this manuscript is a key step to extend the universal computation register of~\cite{pulido-mateoArbitraryQuantumCircuits2024} to a multi-zone trap array based on the Quantum CCD architecture~\cite{winelandExperimentalIssuesCoherent1998a,kielpinskiArchitectureLargescaleIontrap2002}. Here we present a linear trap with zones dedicated to information storage, state readout and computation. We propose a topology based on two storage registers with a computational zone in between, where state readout and cooling will be also executed. To achieve all-to-all connectivity, the computational zone should be capable of performing merging and splitting operations for multi-ion Coulomb crystals as well as swapping operations for qubit pairs. The storage register features an electrode multiplexing approach \cite{holz2DLinearTrap2020} that minimizes the number of signals for operation.

The resulting trap layout is shown in Fig.~\ref{fig: layout}. The rest positions of the ions are marked with blue dots. In order for them to move between these positions, a shuttling operation is required, allowing us to identify the following primitive operations: single-ion transport to move ions between storage slots and along the computational zone, two-ion merging/splitting to prepare crystals for multi-qubit gates, and two-ion swapping to exchange the positions of two ions in a crystal for reordering and all-to-all connectivity. In the following, we describe first the general design process to later relate to the implementation of the primitive operations.

\subsection{Layout considerations}

As the computational zone is a multipurpose structure, it represents the most sensitive aspect of the overall design process and is therefore discussed in more detail. As in previous designs~\cite{carsjensSurfaceelectrodePaulTrap2014, wahnschaffeSingleionMicrowaveNearfield2017, zarantonelloRobustResourceEfficientMicrowave2019a}, the NFQC microwave electrodes are placed in between a set of radio-frequency electrodes of unequal width. Our design hence relies on DC electrodes located outside of the RF and microwave electrodes, effectively dictating the ion-to-control-electrode distance. The computational zone is constructed with 14 outer DC electrodes. To determine their widths, we first defined the maximum and minimum size of the computational zone. In this design, the computational zone size is constrained by the length of the chip. However, as the computational zone and the two storage registers need to be allocated within this length, we need to consider the size requirements of both structures. First, storage registers require sufficiently large electrode sizes to provide enough axial trap depth. Therefore, the maximum computational zone size is constrained to the desired depth of the 16 storage slots. On the other hand, the minimum size of the computational zone is constrained by the length of the microwave meander. A sufficiently long meander lowers undesired residual magnetic fields at the position of gate execution \cite{wahnschaffeSingleionMicrowaveNearfield2017}. Besides the space constraints, to fully determine the computational zone electrode widths, we need to consider the primitives. In other words, we choose an electrode configuration that ensures the correct execution of the primitives. As we will see in further detail in Sec.~\ref{sec:merging and splitting} and Sec.~\ref{sec:swap}, the performance of these operations largely depend on the voltage magnitudes, which are given by the electrode layout.~This means that the resulting excitations after transport are specific to the electrode configuration. For example, a design with reduced ion-to-electrode distance would result in smaller control voltages. Therefore, considering the same available voltage range, the transport primitives would perform better in such a layout. In this design, certain electrode configurations might benefit the performance of one operation but have the opposite effect on another one. Therefore, the electrode widths are chosen to ensure functionality while compromising on the performance of each primitive operation.

\subsection{Modeling and Trap parameters}\label{sec: modelling and trap parameters}

The electrode design and the subsequent calculation of the control voltages is executed with a trap model based on the gapless plane approximation~\cite{wesenbergElectrostaticsSurfaceelectrodeIon2008a}. This modeling approach allows us to use analytical expressions to describe the total trap potential.
\begin{equation}
\label{eq:total potential}
\Phi_{\mathrm{tot}}(r_0)=\Phi_{\mathrm{st}}(r_0)+\Phi_{\mathrm{psd}}(r_0)\quad , 
\end{equation}
where $\Phi_{\mathrm{st}}(r_0)$ is the static potential, obtained by summing up the contributions of each DC electrode. These individual contributions are obtained by multiplying the normalized potential which the respective electrode produces for 1\,V  with the applied voltage. $\Phi_{\mathrm{psd}}(r_0)$ is the pseudopotential produced by the RF electrodes. Here, we assume that the trap is operated with $^9$Be$^+$ ions and that the RF drive frequency is set to $\SI{88.8}{MHz}$ and the RF amplitude to $\SI{63.5}{V}$. Furthermore, we consider that the control voltages are produced by an ideal Digital-to-Analog Converter (DAC) and an amplifier which delivers an overall maximum voltage amplitude of $\pm$ 50\,V, followed by a filtering stage.

\section{Operation Principles\label{sec: trap para}}

To quantify the performance of our processor in the context of ion transport, we first define a set of instructions. These basic building blocks constitute the fundamental steps from which any ion configuration on the chip can be generated. The number and sequence of these instructions determine both the total number of shuttling operations required and, consequently, the time and heating associated with a given reordering task. This approach allows us to efficiently estimate the overall execution time and the accumulated transport-induced heating.

\begin{figure}
\includegraphics[width=\columnwidth,height=0.4\textheight,keepaspectratio]{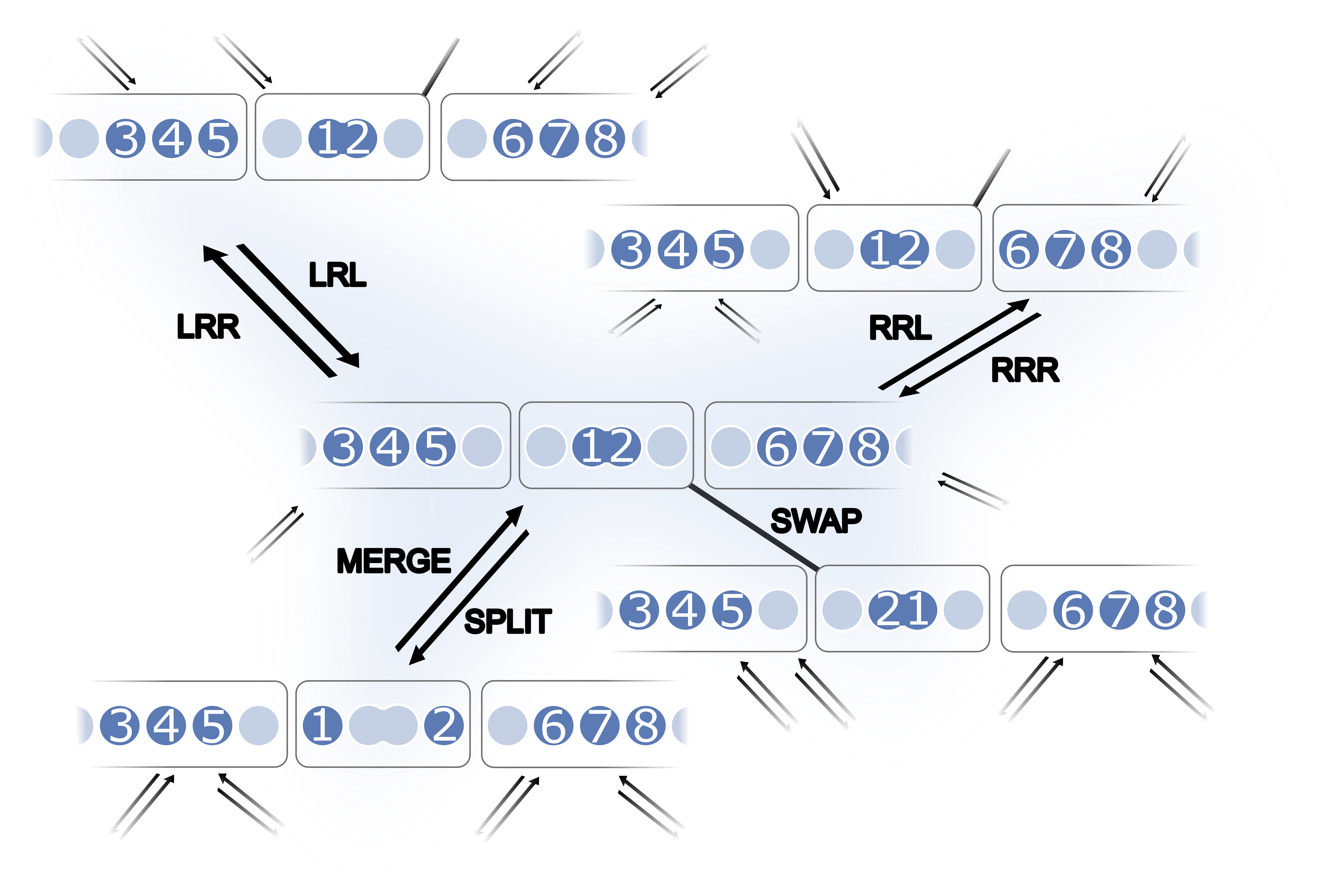}
\caption{Network slice. Nodes represent the different ion positional configurations possible in Fig.~\ref{fig: layout} a), while the edges correspond to instructions connecting them. Each instruction has a cost in time and noise. Only the configuration of the gate zone and the beginnings of the left and right registers are shown. The instructions are labeled only once (\textit{thick black} arrows); the smaller arrows represent the same instructions and indicate possible connections to further configurations.}
\label{fig:network}
\end{figure}

The specific definition of these instructions is largely determined by the chip layout (see Fig.~\ref{fig: layout}). For example, the sixteen potential minima used for storing ions in the left and right registers can only be moved collectively due to the multiplexing scheme~\cite{holz2DLinearTrap2020}. This means that individual ions in these regions cannot be transported independently.
However, four individually controllable resting positions are available in the central gate zone (Fig.~\ref{fig: layout}).
A further example of design-related restriction is that swapping the positions of two ions can only begin once the ions have been transported to the computation zone and merged into a two-ion Coulomb crystal occupying the central trap positions.
Based on these constraints, the set of instructions listed in Table~\ref{tab:instructions} can be identified as one possible realization of basic transport building blocks. Fig.~\ref{fig:network} provides an illustrative excerpt of the resulting configuration network, in which the nodes represent ion configurations and edges correspond to our defined instructions. Here, every edge has a cost associated with it.  

The resting positions available on the chip are indicated in Fig.~\ref{fig: layout} and may or may not be occupied. In principle, the 20 available resting sites would allow for $\frac{20!}{12!}$ possible configurations of eight ions. However, this number is reduced by operational constraints of the trap. 
For instance, conducting a swapping operation can lead to undesired residual static electric fields on the storage registers. This electric field crosstalk originates in the large voltage magnitudes required to execute the swapping operation. To avoid the crosstalk, the first two slots of each register remain unoccupied while the swapping operation is conducted.

The twelve instructions defined here at the compiler level ultimately reduce, on the ion level, to our four primitive operations: the linear transport of a single ion and the three two-ion processes merging, splitting, and swapping, which were defined in the previous Sec.~\ref{sec:Chip Design}. In the following, we characterize these four operations in detail and extract their respective execution costs.

\begin{table}[h]
\caption{Set of instructions to implement basic transport building blocks and their descriptions based on Fig.~\ref{fig: layout}.}
\label{tab:instructions}
\begin{tblr}{
  width=0.9\textwidth,
  colspec={Q[0.2\columnwidth,j,t] Q[0.7\columnwidth,j,t]}
}
\hline\hline
Instruction & Description \\ \hline
RRR         & { Move all ions in the right register one position to the right.}                                                         \\ \hline
RRL         & { Move all ions in the right register one position to the left.}                                                          \\ \hline
GTRL        & { Move the ion in the gate zone from the center to the right.}                                                            \\ \hline
GTRR        & { Move the ion in the gate zone from the right to the center.}                                                            \\ \hline
GTLL        & { Move the ion in the gate zone from the center to the left.}                                                             \\ \hline
GTLR        & { Move the ion in the gate zone from the left to the center.}                                                             \\ \hline
LRR         & { Move all ions in the left register one position to the right.}                                                          \\ \hline
LRL         & { Move all ions in the left register one position to the left.}                                                           \\ \hline
GZ          & { Connect the two center positions in the gate zone for single-ion transport.}                                            \\ \hline
SWAP        & { Swap the position of the ions in the center of the gate zone.}                                                          \\ \hline
MERGE       & { If the left and right sides of the gate zone are occupied, merge the two into one well at the center of the gate zone.} \\ \hline
SPLIT       & { If there are two ions in the center of the trap, separate them into two wells.}                                         \\ \hline\hline
\end{tblr}
\end{table}

\section{motional excitation in single-ion transport} \label{sec:single ion}

In most cases, single-ion transport significantly contributes to the number of transport operations in a reordering process. These single-ion transport operations are necessary to access and operate the storage registers as well as to interface the computational zone with the storage registers. Therefore, it becomes important to characterize the noise produced by each of these operations. Here, we address motional excitation resulting from ion acceleration profiles and fluctuations in the trapping frequency during transport. We assume a perfect execution of the control voltages and neglect the effects of anomalous heating (see Sec.~\ref{sec:anomalous}). There are different methods to minimize the motional excitation due to a driven motion in a rigid harmonic oscillator, i.e.\ without changes in the trapping frequency, as well as in a parametric oscillator, where the trapping frequency is modulated -- for instance, to engineer transport profiles that result in no motional excitation, such as shortcuts to adiabaticity~\cite{guery-odelinShortcutsAdiabaticityConcepts2019} or other methods that depend on the perfect timing of the transport profile~\cite{waltherControllingFastTransport2012,reichleTransportDynamicsSingle2006}. However, to match theoretical predictions, these methods demand highly controlled conditions, which might be beyond our hardware capabilities~\cite{nuschkeAnalysisMotionalHeating2025a}. 
Here, we consider simple hyperbolic tangent profiles; these reduce accelerations at the end and at the beginning of transport, minimizing the effect of inertial forces. Their relation to the energy transfer is given by accelerations as a function of time. 

\begin{equation}
\label{eq: hyperbolc tangent}
       x_0(t) = \frac{L}{2}\  \frac{\tanh \left( N_{\text{val}} \frac{2t - T}{T}\right) + \tanh N_{\text{val}}}{\tanh N_{\text{val}}}
\end{equation}

As shown in Eq.~\eqref{eq: hyperbolc tangent}, for a given transport time $T$ and distance $L=x_\mathrm{final}-x_\mathrm{initial}$, we parameterize the profile using the variable $N_{\text{val}}$. Regardless of $N_{\text{val}}$,  we have $x_0(T/2)=L/2$. The unitless parameter $N_{\text{val}}$ allows us to tune the slope of the transport profile at $T/2$, effectively allowing us to shape the acceleration distribution. While small $N_{\text{val}}$ values converge to a linear transport profile with impulsive initial accelerations, larger ones converge to near-zero initial acceleration with transport profiles approaching a step function.

In this work, we aim to characterize the energy transfer within two specific regions: the \textit{noisy regime}, defined here as $\bar{n}>1$, and the \textit{near-adiabatic} regime ($\bar{n}<0.1$). To calculate the average phonon number $\bar{n}$, one may begin with the time-dependent Schrödinger equation for an ion confined in a potential well that undergoes acceleration,
\begin{equation}
\begin{aligned}\label{schroedinger}
    i\hbar\frac{\partial\psi(s,t)}{\partial t}=
    &-\frac{\hbar^2}{2m}\frac{\partial^2\psi(s,t)}{\partial s^2}+\frac{1}{2}m\omega^2(t)s^2\psi(s,t)\\
    &-f(t)s\psi(s,t)
\end{aligned}  
\end{equation}
where, in our case, $s = x - x_0(t)$ denotes the displacement relative to the moving trap minimum $x_0(t)$, and $f(t) = -m\ddot{x}_0(t)$ represents the effective driving force induced by the trap acceleration.

By constructing the propagator of the forced parametric harmonic oscillator, one can evaluate the transition probabilities from the motional ground state to a given excited Fock state. Employing a generating-function ansatz, these transition probabilities can be cast into a closed-form expression, from which a compact analytical formula for the average phonon number $\bar{n}$ follows (see Appendix~\ref{sec:appendix_propagator}). The resulting expression naturally separates into two distinct contributions, reflecting the influence of inertial forcing on a parametric oscillator and the effect of changes in frequency in a resting frame~\cite{husimiMiscellaneaElementaryQuantum1953}. These are given by Eq.~\eqref{eq:motional excitation}, where $\eta$ represents the contribution of the inertial forcing, remarkably obtained by means of the classical solutions of the forced parametric oscillator, with the classical energy gain $E_{kin}/\hbar\omega_0$.  In this case, the gain in kinetic energy originates from the resonance condition between the harmonic oscillator\footnote{Considering a harmonic potential with the resonance frequency $\omega_0$.} and the Fourier component of the transport acceleration profile corresponding to its frequency~\cite{reichleTransportDynamicsSingle2006}. 
\begin{equation}
\label{eq:motional excitation}
\langle n_0(T) \rangle = \eta(T,0) + \frac{1}{2} \left( Q(T,0) - 1 \right)
\end{equation}

$Q$, on the other hand, can be understood as the contribution to the energy gain from the frequency variation alone, independent of any displacement. This occurs because when the frequency of the harmonic oscillator is modulated, as is typical for the parametric oscillator, the original ground state may appear as an excited state within the new frequency regime. This results in a mixing of states that can be quantified using the Bogoliubov transformation. The full expression can be seen in Appendix~\ref{sec:appendix_propagator}, where the dependence on the classical solutions of the parametric oscillator can be noticed.
For our primitive operations with a modulated trapping frequency, $\eta$ exceeds $\frac{1}{2}\left(Q(T,0)-1\right)$, the excitation generated out of the motional ground state by the frequency modulation alone, by orders of magnitude on the time scales of interest (see Appendix~\ref{sec:appendix_bogoliubov}). The latter is therefore neglected in the following.

Due to the fact that the individual contributions to phonon excitation could be generated from classical contributions, we can always refer to classical numerical simulations to determine the heating in the following\footnote{Strictly speaking, the analytical derivation of this quantum--classical correspondence only holds for harmonic potentials. However, \textsc{QuTiP} simulations of the anharmonic model show that the corrections are negligible for our final results in Fig.~\ref{fig:all-to-all}~b); see Appendix~\ref{appendixC}.}.

\begin{figure}
\includegraphics[width=\columnwidth]{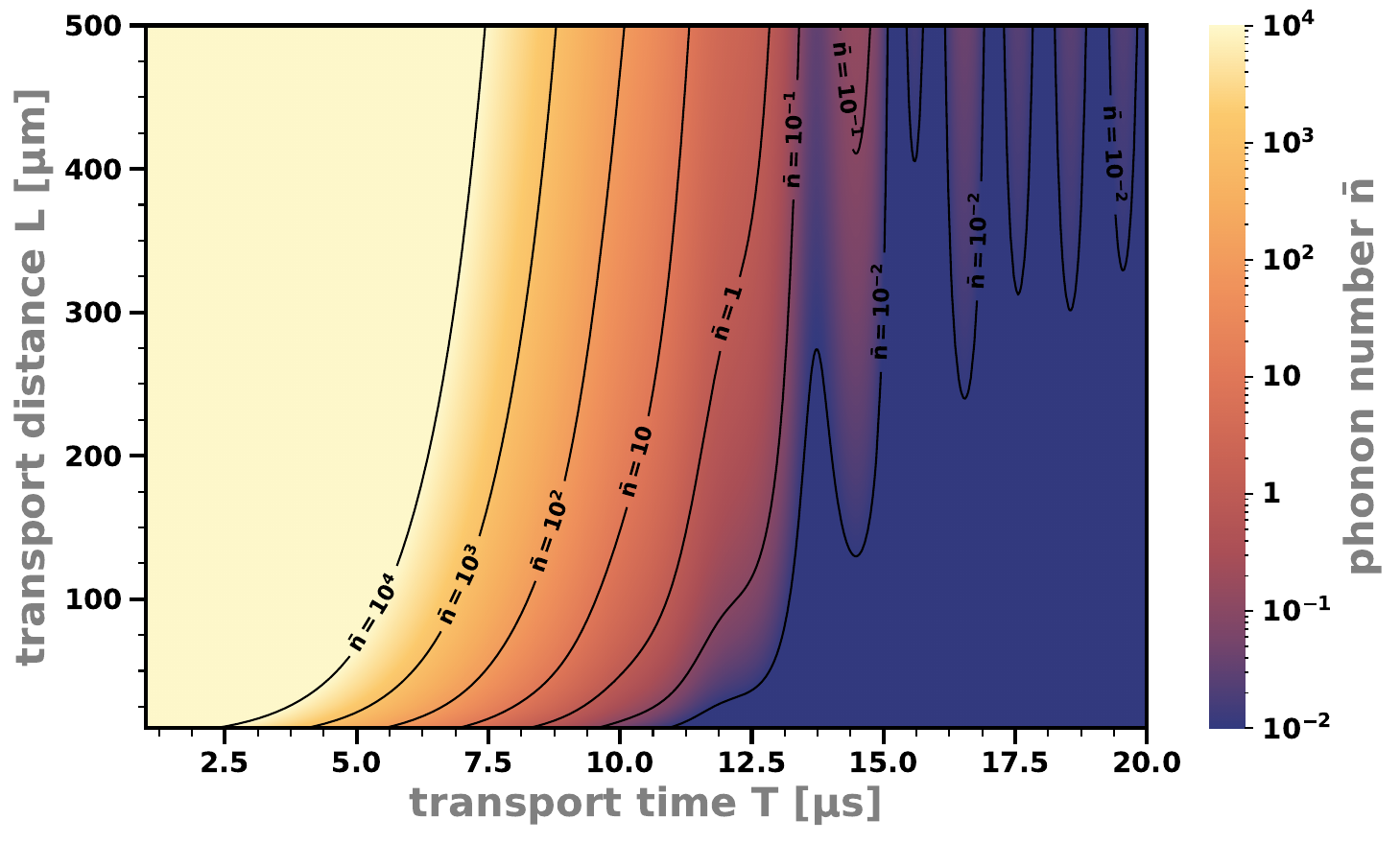}
\caption{Energy transfer for a single transported ion as a function of distance and transport time. A hyperbolic tangent with $N_{\text{val}}=5$ is selected.}
\label{fig:transport_distances}
\end{figure} 

In the context of single-ion transport, we here consider operations with a constant trapping frequency of 1\,MHz during transport. The physical distance that the ions are to go through can be observed in Fig.~\ref{fig: layout} as the distance between the resting positions. Since the distance between registers is the same and the trap is symmetric along the transport axes, we only have to consider three different transport distances. We model the potential using the form of a driven harmonic oscillator, expressed as $U(x,t)=\frac{1}{2}m\omega_x^2(x-x_0(t))^2$, where $x_0(t)$ denotes the time-dependent position of the potential minimum\footnote{In the following sections, we will not only consider these \textit{idealized potentials} but also take the electrode geometry and calculated control voltages into account. This increases the computational effort; however, it can lead to very similar outcomes.}. The force that is applied to the particle due to the change in minimum position might be expressed as:
\begin{equation}
    m\ddot{x}=-\frac{\partial U}{\partial x} = m\omega_x^2(x_0(t)-x)\quad.
\label{eq: simple harmonic oscillator motion}
\end{equation}

To determine the average phonon number after transport, we first obtain the classical solution by solving the equation of motion as given in Eq.~\eqref{eq: simple harmonic oscillator motion}\footnote{This inhomogeneous differential equation results in the summation of the homogeneous solution and a particular solution, where the latter depends on $\ddot{x}_0$ and $\omega_x$. Since we start at rest, the homogeneous solution is trivial; therefore, the motion is given solely by the particular solution.}, starting from rest. We then use Eq.~\eqref{eq:motional excitation} to find $\bar{n}$.

For example in Fig.~\ref{fig:transport_distances}, through the parametrization of Eq.~\eqref{eq: hyperbolc tangent}, we calculate $\bar{n}$ for different transport times and distances. It can be observed that the phonon number shows higher sensitivity to transport time than to distance. In this case the \textit{near-adiabatic} regime is reached within $20\,\mu$s regardless of the transport distance.
In addition, we observed that, for single-ion transport, considering the trapping potentials created by calculated control voltages and the electrode geometry~\cite{nuschkeAnalysisMotionalHeating2025a} results in a similar outcome for the time scales of interest. Therefore, for single-ion transport we will rely on idealized potentials in the following.

\section{merging and splitting} \label{sec:merging and splitting}
\subsection{Geometrical considerations for two-ion merging}
\begin{figure}
    \centering
    \includegraphics[width=\columnwidth]{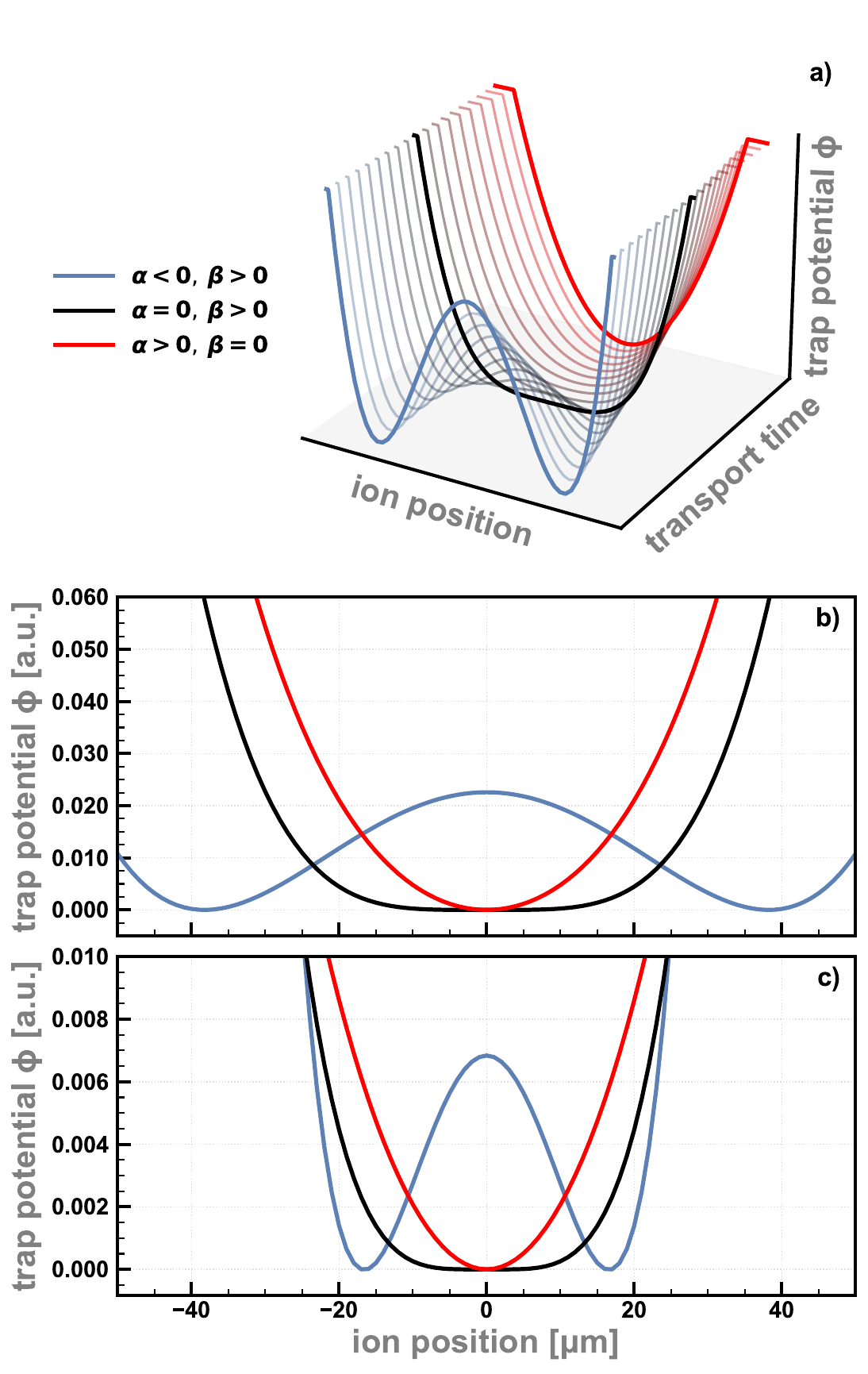}
    \caption{The three stages for merging or splitting a two-ion crystal. \textit{a}) legend and three dimensional overview. The process is here exemplified by either \textit{b}) maintaining a constant distance between the ions while varying the trapping frequency, or \textit{c}) keeping the trapping frequency constant while adjusting the ion distance. The corresponding frequency trajectories with respect to the ion distance are shown in Fig.~\ref{fig:ideal}.}
    \label{fig:example_merge}
\end{figure}

As seen in~\cite{homeElectrodeConfigurationsFast2006a}, bringing two ions from two different wells into a single well requires the formation and precise control of quadrupole and octupole potentials. We first identify the required characteristics of the theoretical potential that needs to be produced by the DC electrodes and the control voltages applied to them. Considering a one-dimensional potential and expanding up to fourth order at the origin, we obtain:
\begin{equation}
\label{eq:quartic potential}
\phi(x)=\alpha x^2+\gamma x^3+ \beta x^4 \quad ,
\end{equation}
where the point of expansion is taken to be a local extremum implying a vanishing linear term. Moreover, as offsets do not contribute to the motion, these are neglected, and since the potential is symmetric around the center, we assume $\gamma = 0$. Furthermore, we declare an auxiliary function that accounts for the Coulomb repulsion:
\begin{equation}
\label{eq:coulomb potential}
C_P(x_1,x_2)=\frac{q}{4\pi \epsilon_0(x_1-x_2)}\quad .
\end{equation}
The total potential energy is then given by:
\begin{equation}
\label{eq:quartic potential  and couloumb}
\Phi=\phi(x_1)+\phi(x_2)+ C_P(x_1,x_2) \quad .
\end{equation}
As seen similarly in~\cite{homeElectrodeConfigurationsFast2006a} we can divide the merging process into three different stages (with reference to Fig.~\ref{fig:example_merge}):
\begin{enumerate}
\label{three stages}
  \item \textbf{Two separate wells} (\textit{navy blue}): the ions are sitting close together in two different wells with $\alpha<0$.
  \item \textbf{Critical point} (\textit{black}): the ions are being held in place by the Coulomb repulsion and a purely quartic potential, with $\alpha=0$.
  \item \textbf{Single harmonic well} (\textit{red}): the ions are in a single quadratic potential well with $\alpha>0$ and $\beta=0$.
\end{enumerate}
A useful quantity to consider is the axial Center Of Mass (COM) mode frequency $\omega_\mathrm{COM}$. An obvious choice would be to try to obtain a constant $\omega_\mathrm{COM}$ throughout this process~\cite{savasFastSeparationTrapped2026}. However, this imposes a high constraint on $\beta$ (see Eq.~\ref{eq:curvature global merging}), which is difficult to achieve with most setups. This problem can be addressed by decreasing the trapping frequency around the \textbf{critical point}~\cite{kaufmannScalableQuantumProcessor2017}. However, as the trap frequency decreases, the system becomes more sensitive to motional excitation during transport, ultimately limiting the total transport speed~\cite{nuschkeAnalysisMotionalHeating2025a}. 

To what extent this strategy needs to be applied depends on the specific electrode layout. The trap frequency achievable at the \textbf{critical point} for a given limit on trap voltages depends on the strength of the quartic term. While this term can be maximized in the design process by adjusting the electrode width, it makes the resulting geometry less suitable (i.e., increases the maximum voltage required) for other processes such as swapping operations (see Sec.~\ref{sec: Geometrical considerations for two ion swapping}). The design used in this work is therefore a compromise that addresses the requirements of both swapping and merging/splitting operations.

Another way of looking at it is the highest required voltage to obtain a given quartic term as a function of the trap geometry. We find that in our compromise design, this voltage is more sensitive to the targeted value of the quartic term. However, the highest achievable trap frequency within our voltage limits is only $\approx\SI{40}{KHz}$ smaller than in a corresponding geometry optimized for merging and splitting only. As discussed in Sec.~\ref{sec:theoretical evaluation}, this difference hardly affects the overall transport performance. 

\subsection{System constraints} \label{sec: system constraints}
One approach to calculate the control voltages is to constrain the potential derivatives up to 4th order at the center of the merging and splitting zone~\cite{kaufmannScalableQuantumProcessor2017,kaufmannDynamicsControlFast2014}. However, the positions of the ions are then inferred from that local expansion of the potential. The further the obtained ion positions are away from the center, the more this method becomes sensitive to deviations between the real potential and the expanded potential. While this method was successfully demonstrated~\cite{rusterExperimentalRealizationFast2014a}, this may necessitate a two-step process, where the ion positions are updated based on the full potential once a first approximation has been found based on the expanded potential~\cite{negnevitskyFeedbackstabilisedQuantumStates2018}.

Therefore, we choose a different approach here; instead of constraining the potential at the \textit{one} central position, we directly constrain the \textit{two} positions of the ions, obtaining certainty within the simulations about their locations and frequencies. In this case, the calculation of potential derivatives higher than second order is no longer required, since we expand around \textit{two} positions. This can be advantageous for modelling traps using non-analytic methods to calculate basis functions~\cite{contaToolchainShuttlingTrappedion2026}, since these require numerical derivatives. We will refer to our approach as the \textit{two-point constraint} method for ion pairs. 

Next, we present the constraints that the static potential $\Phi_{\mathrm{st}}$ must satisfy to realize the three stages (see Sec.~\ref{three stages}) of a merging or splitting operation. First, the two ions need to align with the RF null line. To achieve this, the radial electrostatic fields, $E^\mathrm{st}(r_0)$, must cancel at the RF null positions. This is accomplished when the condition $E^\mathrm{st}_y(r_i)\doteq E^{\mathrm{st}}_z(r_i)\doteq0|_{i\in [1,2]}$ is satisfied, where $r_1$ and $r_2$ are the ion positions. Additionally, to align the motion of the axial modes with the trap axis, we constrain the cross derivatives to zero ($\partial E^{\mathrm{st}}_x(r_i)/\partial y\doteq\partial E^{\mathrm{st}}_x(r_i)/\partial z\doteq0|_{i\in [1,2]}$). This decouples the axial motion from the radial components.

By symmetry, the location of the two ions is equidistant from the origin, $x_1=-d/2$ and $x_2=+d/2$, with $d$ the distance between the ions. Let $E_\mathrm{C}$ be the electric field that each of the ions feels due to the presence of the other ion:
\begin{equation}
\label{eq:coulumb repulsion}
  E_\mathrm{C}(d)=\frac{q}{4\pi \epsilon_0 d^2} \quad .
\end{equation}
The voltage vector that ensures the desired position of the ions can be obtained by solving $\frac{\partial\Phi_{}}{\partial x_1 }=0$ and $\frac{\partial\Phi_{}}{\partial x_2 }=0$, or equivalently:
\begin{equation}
\label{eq:position merging}
  \frac{\partial\Phi_{\mathrm{st}}}{\partial x }\Bigg|_{x=\pm d/2}=\pm E_\mathrm{C}(d) \quad .
\end{equation}
$\omega_\mathrm{COM}$ of each ion is only given by the local trap potential curvature at the equilibrium positions of each ion, as the Coulomb repulsion is balanced by the first order term of the potential expansion around each ion's position.

Let us go back to Eq.~\eqref{eq:quartic potential} to obtain an intuitive picture. The local potential curvature and COM frequency at each equilibrium position is then given by
\begin{equation}
\label{eq:curvature global merging}
    \begin{cases}
      \frac{\partial^2\phi(x_1)}{\partial x_1^2 }\Big|_{x_1=-d/2}=2 \alpha +3 \beta d^2=\frac{m}{q}\omega^2_\mathrm{COM}\\
      \frac{\partial^2\phi(x_2)}{\partial x_2^2}\Big|_{x_2=d/2}=2 \alpha +3 \beta d^2=\frac{m}{q}\omega^2_\mathrm{COM}
    \end{cases} \quad ,
\end{equation}
since we set $\gamma = 0$. We could prescribe the COM frequency by manually tuning $\alpha$ and $\beta$ to obtain the desired $\omega_\mathrm{COM}$ value. Unless there is a specific advantage to obtaining certain values of $\alpha$ and $\beta$, however, it is better to directly select a target frequency and let the minimization algorithm adjust the coefficients accordingly.

We continue with the development of an intuitive picture by examining the three different stages of merging and splitting mentioned in Sec.~\ref{three stages}. A merging process starts with both ions in \textbf{two separate wells}, parted by a small potential barrier. If the initial interatomic distance $d_\mathrm{in}$ is large enough, the ion positions approach the potential minima, with Eq. \eqref{eq:coulumb repulsion} $\approx$ 0. 

At the \textbf{critical point}, we have $\alpha=0$ and the potential is only defined by $\beta$. This allows us to obtain a direct relation between $d$ and $\beta$: 
$$\beta(d)=\frac{q}{2\pi\epsilon_0 d^5} \quad .$$
The frequency is then given by
\begin{equation}
\label{eq:curvature_critical}
       \frac{\partial^2\Phi_{\mathrm{st}}}{\partial x^2 }\Bigg|_{x=\pm d/2}=\frac{3q}{2\pi \epsilon_0 d^3}\quad .
\end{equation}

Therefore, if Eq.~\eqref{eq:position merging} and Eq.~\eqref{eq:curvature_critical} hold, the real potential will approximate to a purely quartic potential, hosting the two ions at a constrained critical distance $d_\mathrm{crit}$ with frequency $\omega_\mathrm{crit}$.

This also suggests that we are able to tune the critical frequency $\omega_\mathrm{crit}$, during the merging process, by reaching the critical distance with different values of $d_\mathrm{crit}$. We can therefore raise the minimum frequency by letting the ions sit closer together at the \textbf{critical point}, which is analogous to having a high quartic term at the critical distance\cite{nizamaniOptimumElectrodeConfigurations2012,kaufmannFastIonSwapping2017}. 

At the final stage, when the two ions are in a \textbf{single harmonic well}, we have $\beta = 0$. Here the potential is defined only by $\alpha$, with a relation to $d$ of the form:
$$\alpha(d)=\frac{q}{4\pi\epsilon_0 d^3}\quad.$$
The frequency is now determined by:

\begin{equation}
\label{eq:curvature_one_Well}
       \frac{\partial^2\Phi_{\rm{st}}}{\partial x^2 }\Bigg|_{x=\pm d/2}=\frac{q}{2\pi \epsilon_0 d^3}\quad.
\end{equation}

If the constraints in Eq.~\eqref{eq:curvature_one_Well} and Eq. \eqref{eq:position merging} hold, the real potential will approximate to a purely quadratic potential with an equilibrium distance, $d_{\mathrm{final}}$ and frequency $\omega_{\mathrm{final}}$.

\begin{figure}
\includegraphics[width=\columnwidth]{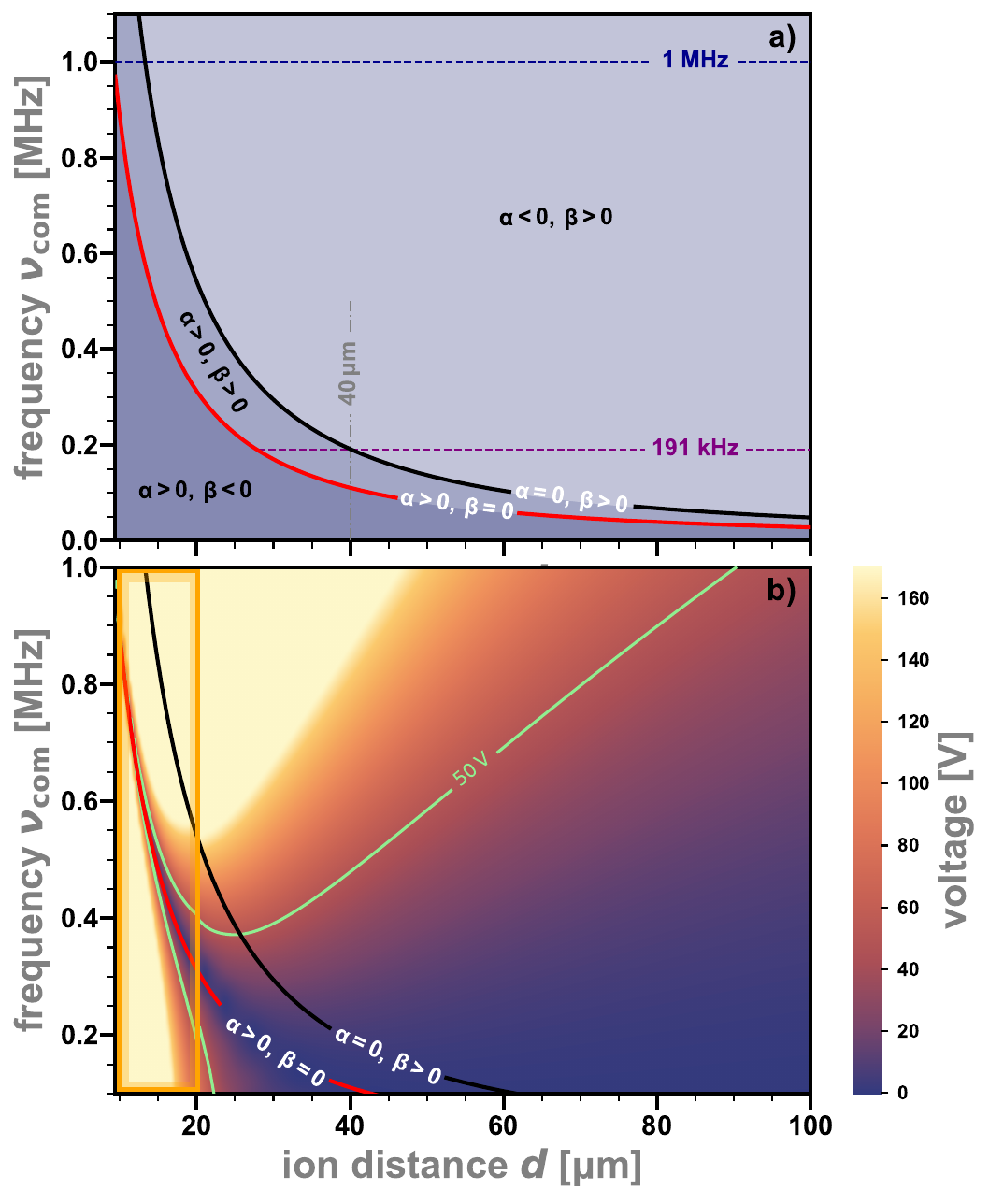}
    \caption{Equilibrium distance vs COM frequency for different trapping potential shapes. Purely quadratic potential (\textit{red line}), purely quartic potential (\textit{black line}). \textit{a}) mixed potential regions and examples of constant frequency trajectory at $\SI{1}{MHz}$ (\textit{blue line}) and $\SI{191}{KHz}$ (\textit{purple}). A second example for constant distance (\textit{gray}). \textit{b}) heatmap of maximum Voltages necessary. Only values below $50\,$V (\textit{green contour line}) are within the hardware specifications. The orange frame marks an area of increased resolution in frequency space ($10\,$kHz compared to $50\,$kHz). A cubic interpolation is further used in both cases.}
\label{fig:ideal}
\end{figure}

The frequency dependence on the equilibrium distance we developed during the three stages (see Sec.~\ref{three stages}) of the merging process is summarized in Fig.~\ref{fig:ideal}. The case of a purely quadratic potential is shown in the \textit{red curve} and the case of a purely quartic potential in the \textit{black} curve using Eq.~\eqref{eq:curvature_critical} and Eq.~\eqref{eq:curvature_one_Well}, respectively. As we constrain the distance between ions and the trapping frequency, the exact shape of the real potential might not seem important. However, to identify the combinations of $d$ and $\omega_{\rm{COM}}$ that will result in unfavorably high voltages (see Fig.~\ref{fig:ideal} \textit{bottom}), we need to have a potential approximation in terms of these parameters. This completes the development of the intuitive picture and allows us to choose the parameters in terms of distance and frequency in the following.\\\\

For a given desired mode frequency, we can identify one value of $d$ in each case, corresponding to a purely harmonic or a purely quartic potential (see \textit{blue line}); for instance, a frequency of 
$\SI{191}{kHz}$ (see \textit{magenta line}) results in the potential shapes observed in Fig.~\ref{fig:example_merge} b). In the latter case, the required voltages are relatively low due to the reduced mode frequency. However, a target mode frequency of 1\,MHz, (\textit{blue line}), would necessitate voltages higher than what our hardware setup allows. This is a direct consequence of the large octupole term required to maintain the ions in two different wells and to raise $\omega_{\mathrm{crit}}$ in our particular layout.

In the next section, we present a frequency trajectory that minimizes $\bar{n}$ and can be implemented within our hardware specifications (see Fig.~\ref{fig:frequency_trajectories}~b)). This frequency trajectory, $\nu_\mathrm{COM}(d(t))$ (hereafter referred to as $\nu_\mathrm{COM}(t)$), is obtained through re-parametrization via the transport profile $d(t)$ and is derived by interpolating the frequency constraints presented above.

\subsection{Theoretical evaluation} \label{sec:theoretical evaluation}

\begin{figure}
    \centering
    \includegraphics[width=\columnwidth]{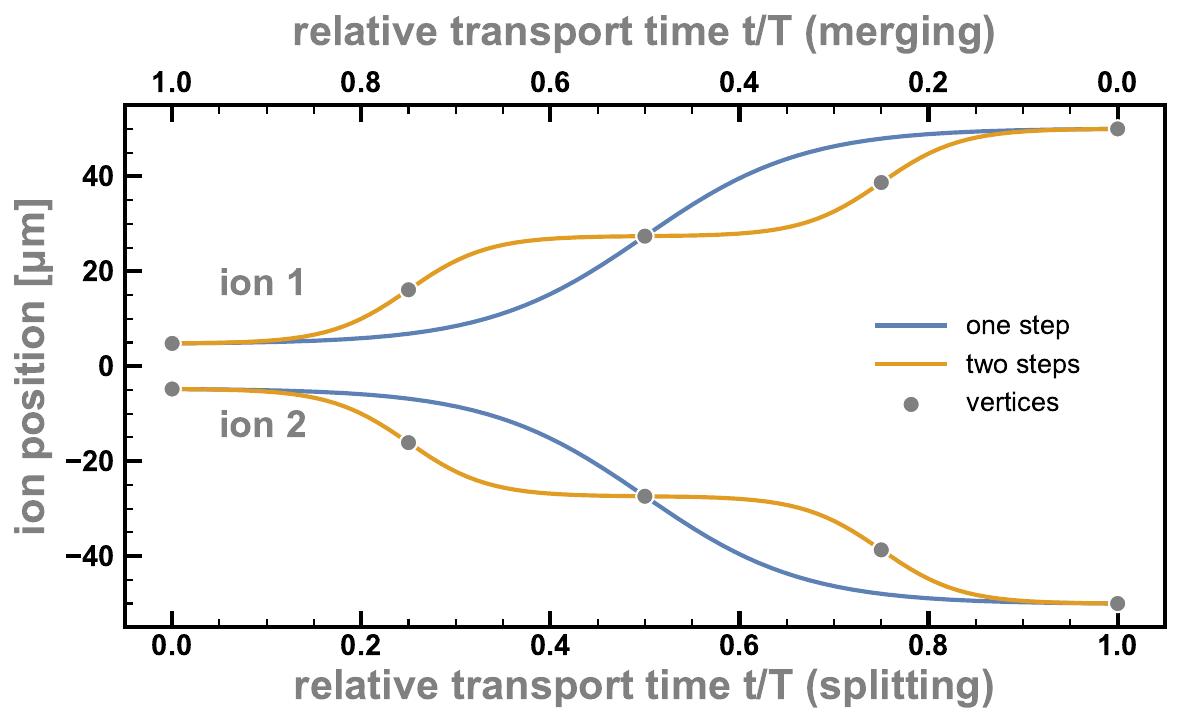}
    \caption{Ion positions during splitting or merging for different transport profiles.
    \textit{Blue}: single hyperbolic tangent with $N_{\mathrm{val}}=3$ applied for operations featuring a symmetric frequency trajectories $\nu_{\mathrm{com}}(t)$. \textit{Yellow}: two sequentially applied hyperbolic tangent profiles, both with $N_{\mathrm{val}}=3$ are used for operations featuring an asymmetric frequency trajectory $\nu_{\mathrm{com}}(t)$.
    Depending on whether the profile has two steps or one, it features 3 or 5 vertices, i.e. points insensitive to $N_{\mathrm{val}}$.}
    \label{fig:transport profiles}
\end{figure}

Based on the developed intuitive understanding, we can characterize the maximum $\omega_{\mathrm{crit}}$ achievable within our hardware constraints by identifying the minimum distance at which a purely quartic potential can be realized with voltages below $50\,\mathrm{V}$ (see Fig.~\ref{fig:ideal}~b). Next, we need to construct the frequency interpolation in between the three stages of merging. Apart from limiting the maximum $\omega_{\mathrm{crit}}$, the voltage output constrains the interpolation function as well. Besides completing the two-ion merging, we want to minimize the associated phonon excitation. Looking back at the general case of single-ion transport (see Sec.~\ref{sec:single ion}), it is known that the transport profile and the trapping frequency influence the phonon excitation $\bar{n}$. Thus, now in the case of two-ion merging a motional heating analysis is performed for different transport profiles $d(t)$ (see Sec.~\ref{sec: trap para}), shown in Fig.~\ref{fig:transport profiles}, and for different frequency interpolations.\\\\

First, we analyze the sensitivity of the motional excitation, $\bar{n}$, to the transport profile and to the frequency interpolation when the critical frequency, $\omega_{\mathrm{crit}}$, and therefore $d_{\mathrm{crit}}$, are fixed. For convenience, this analysis can be conducted using a frequency interpolation symmetric in time around the minimum frequency first, as the parametrization of such interpolations requires a smaller number of parameters than in the general case.
Afterwards, using the results of this sensitivity analysis, we identify the dominant contribution to $\bar{n}$ when changing $\omega_{\mathrm{crit}}$. We will see, however, that manipulating $\omega_{\mathrm{crit}}$,  cannot be achieved using symmetric frequency interpolations within the available voltage range and the identical total distance $d_{\mathrm{in}}$; we therefore proceed with asymmetric interpolations. In general, for obvious reasons the same frequency trajectory cannot be used on two different $\omega_{\mathrm{crit}}$ values.

Table~\ref{table: sym vs asym} shows the general constraints we impose on the characteristic parameters for transport profiles and frequency trajectories used in this section. The starting point for these parameters is the desired $\nu_\mathrm{in}=\nu_\mathrm{final}=1\,\mathrm{MHz}$. The distances are the result of realizing these frequencies while staying within the experimental voltage limits. \\\\

\begin{figure}
    \centering
    \includegraphics[width=\columnwidth]{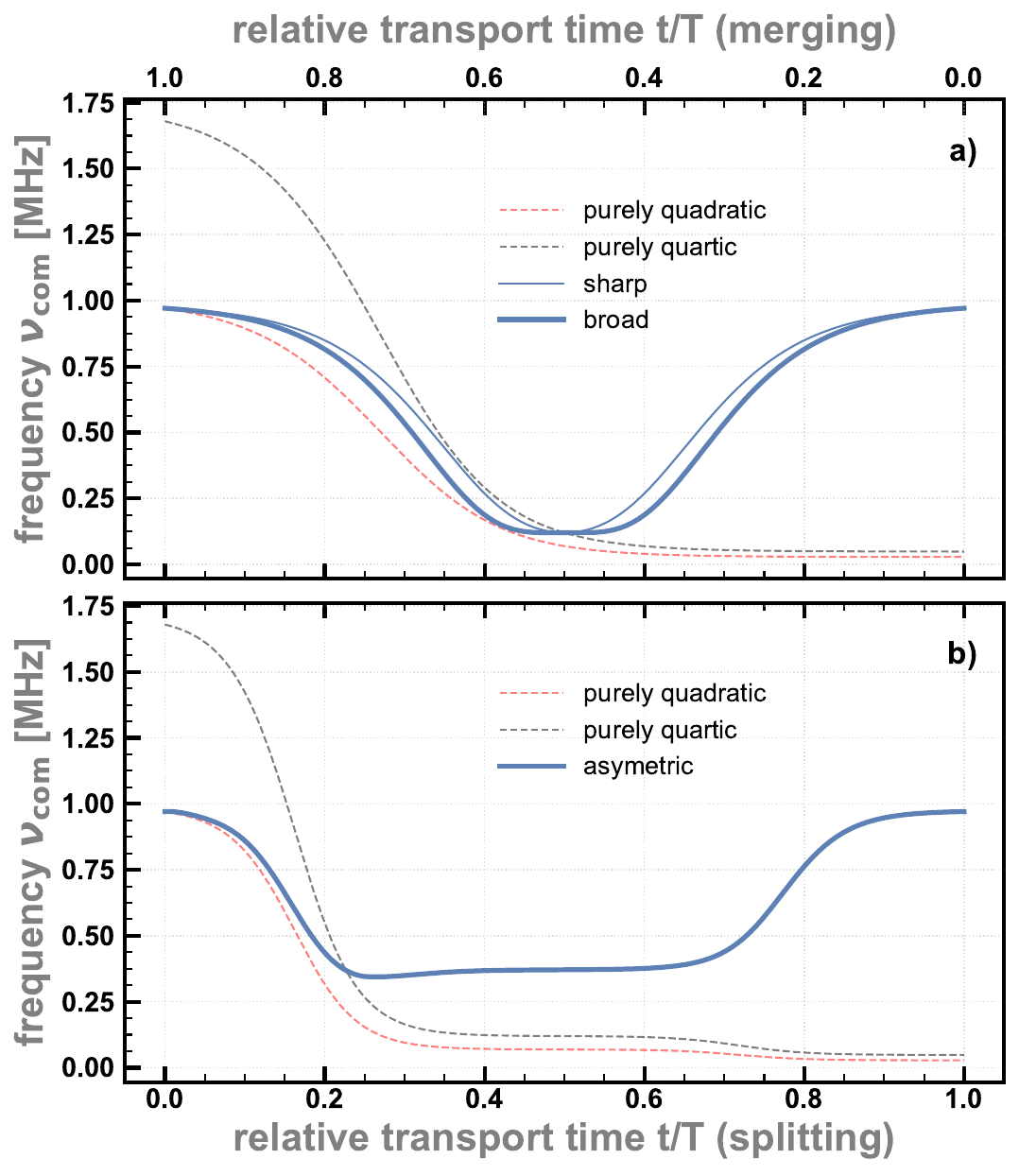}
    \caption{Frequency trajectories $\nu_{\mathrm{com}}(t)$ for two different transport schemes: \textit{a}) symmetric and \textit{b}) asymmetric, in the context of splitting and merging. In \textit{a}), the “one-step” transport profile, and in \textit{b}), the “two-step” transport profile from Fig.~\ref{fig:transport profiles} are used. The red and black lines represent the time-dependent ion distances for purely quadratic and quartic potentials, respectively. Intersections with the transport schemes (\textit{blue lines}) indicate the times at which the corresponding potential characteristics are realized. }
    \label{fig:frequency_trajectories}
\end{figure}

\begin{table}
  \caption{Constraints of the frequency trajectories and transport profiles}
  \label{table: sym vs asym}
  \begin{tblr}{
			width=\columnwidth,
			colspec={Q[1,l,t] | Q[1,c,t] | Q[1,c,t]}
  }
    \hline\hline
    \SetCell[c=3]{c,t} Trajectory features & & \\ \hline
    $\omega(t)$                            & symmetric                                    & asymmetric                      \\ \hline\hline
    $d_{\rm in}$                           & \SetCell[c=2]{c,t} $100\,\mu\mathrm{m}$      &                                 \\ \hline
    $d_{\rm crit}$                         & $50\,\mu\mathrm{m}$                          & $47.7-27\,\mu\mathrm{m}$        \\ \hline
    $d_{\rm final}$                        & \SetCell[c=2]{c,t} $9.4\,\mu\mathrm{m}$      &                                 \\ \hline
    $\nu_{\rm crit}$                       & $\approx 120\,\mathrm{kHz}$                  & $\approx 190-350\,\mathrm{kHz}$ \\ \hline
    $\nu_{\rm in}=\nu_{\rm final}$         & \SetCell[c=2]{c,t} $\approx 1\,\mathrm{MHz}$ & \\
    \hline\hline
  \end{tblr}
\end{table}

We now discuss the heating analysis in more detail. For a given transport profile and frequency interpolation, we present the differential equation describing the merging and splitting procedure.
As a first approximation, we will consider the theoretical potential shown in Eq.~\eqref{eq:quartic potential  and couloumb}. Diagonalizing its Hessian, $H(\Phi)$, allows us to first express the potential with relative coordinates and in terms of the mode frequencies 
$\omega_{\mathrm{com}}(t)$ and $\omega_{\mathrm{str}}(t)$. Here $\omega_{\mathrm{com}}(t)$ depends on the prescribed time-frequency dependency (see Fig.~\ref{fig:frequency_trajectories}). Since 
\begin{equation}
    \label{eq: STRvsCOM}
    \omega_{\mathrm{str}}(t)^2=\omega_{\mathrm{com}}(t)^2+\frac{q^2}{\pi \epsilon_0 m d(t)^3}\quad,
\end{equation} the stretch mode frequency depends on the COM frequency and transport profile as well~\cite{homeElectrodeConfigurationsFast2006a}. After rewriting the potential in terms of the mode frequencies and the global coordinates $x_1$ and $ x_2$, we obtain the following coupled system of equations of motion:

\begin{equation}
\label{eq:equations of motion}
\left\{
\begin{aligned}
\ddot{x}_1 &= -\omega(t)_{\mathrm{com}}^2 \left( x_1 - \frac{d(t)}{2} \right) \\
           &\quad - \left( \frac{\omega(t)_{\mathrm{str}}^2 - \omega(t)_{\mathrm{com}}^2}{2} \right)\left( x_1 - x_2 -d(t) \right) \\
\ddot{x}_2 &= -\omega(t)_{\mathrm{com}}^2 \left( x_2 + \frac{d(t)}{2} \right) \\
           &\quad + \left( \frac{\omega(t)_{\mathrm{str}}^2 - \omega(t)_{\mathrm{com}}^2}{2} \right)\left( x_1 - x_2-d(t) \right)\, .
\end{aligned}
\right.
\end{equation}
We solve these using \textsc{NDsolve} in \textsc{Mathematica} and calculate $\bar{n}$ using Eq.~\eqref{eq:motional excitation}.

\begin{figure}
\centering
\includegraphics[width=\columnwidth]{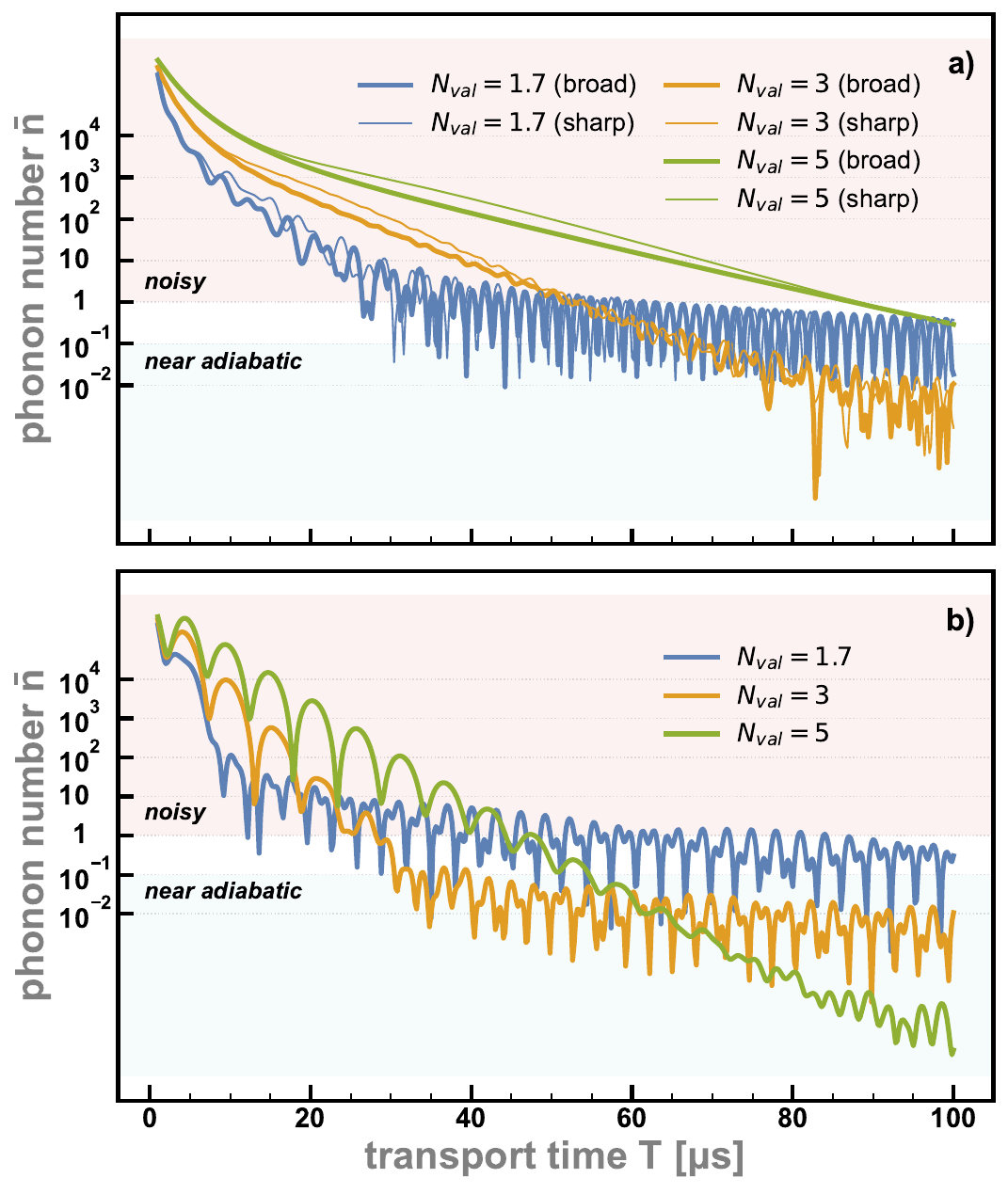}
    \caption{Average phonon excitation vs. transport time in the context of two-ion splitting; the excitation in the case of merging is comparable, i.e., negligibly different. The \textit{noisy} regime for high levels and the \textit{near adiabatic} regime for low levels of phonon excitation are shaded in \textit{red} and \textit{blue}, respectively. The results consider idealized potentials and do not take the electrode geometry or voltage interpolation into account; for this, see Fig.~\ref{fig:comparsion_interpol}. The energy transfer for \textit{a}) symmetric and \textit{b}) asymmetric frequency dips (the different frequency dips are presented in Fig.~\ref{fig:frequency_trajectories}) are shown. For each frequency dip, three hyperbolic transport profiles with different $N_{\text{val}}$ parameters were selected. Additionally, \textit{a}) shows the sensitivity to the frequency interpolation by comparing the \textit{broad} to the \textit{sharp} profile from Fig.~\ref{fig:frequency_trajectories} a). Moreover, for transport times $T<100\mu$s, the curve corresponding to $N_{\text{val}}=5$ in \textit{b}) reaches smaller excitations than the curve for $N_{\text{val}}=1.7,3$. In \textit{a}), this is only the case for $T>100\mu$s. This behaviour scales with $\frac{1}{\omega_0}$.}    \label{fig:comparison_asym_heat}
\end{figure}

The evaluation of the symmetric frequency trajectory is shown in Fig.~\ref{fig:frequency_trajectories}~a). It is observed that $\bar{n}$ shows a stronger sensitivity to the utilized transport profiles than to the given shape of the frequency trajectory (see Fig.~\ref{fig:comparison_asym_heat}~a)). This becomes relevant when examining the sensitivity with respect to $\omega_{\rm{crit}}$, as for every value of $\omega_{\rm{crit}}$ we will obtain a different frequency interpolation. In such case, we associate large changes in the resulting phonon number $\bar{n}$ to the variations of $\omega_{\rm{crit}}$ and not to the corresponding frequency trajectory.

We now focus on the influence of the transport profile type (see Eq. \eqref{eq: hyperbolc tangent}), which was chosen to keep $d_{\mathrm{crit}}(t)$ insensitive to a slope parameter ($N_{\mathrm{val}}$)\footnote{This condition is valid if $d_{\mathrm{crit}}(t)$ is located at a vertex of the transport profile. For example to realize the two-ion crystal splitting/merging using the asymmetric frequency trajectory in Fig.~\ref{fig:frequency_trajectories}~b), $d_{\mathrm{crit}}(t)$ must be located at the second/forth vertex of the two step transport profile in Fig.~\ref{fig:transport profiles}.}. This allows us to analyze the behavior as a function of this parameter (see Fig.~\ref{fig:comparison_asym_heat}). Hyperbolic tangents with steeper slopes at $d_{\mathrm{crit}}$, i.e., higher $N_{\mathrm{val}}$, consistently result in lower $\bar{n}$ for large transport times. However, for all $N_{\mathrm{val}}$ values, the phonon excitation initially decreases rapidly before transitioning to a slower-decay regime. For the steeper profiles, this transition occurs at longer transport times than for profiles with lower $N_{\mathrm{val}}$ values.~\cite{nuschkeAnalysisMotionalHeating2025a}. This behavior is characteristic of hyperbolic tangent profiles, irrespective of $\omega_{\mathrm{crit}}$.
This makes the selection of $N_{\mathrm{val}}$ not trivial for a \textit{noisy regime}, since for a target transport time a certain slope results in a lower $\bar{n}$ value.
This suggests that under certain conditions, a frequency trajectory featuring a smaller $\omega_{\mathrm{crit}}$ can achieve a comparable performance to one with a larger $\omega_{\mathrm{crit}}$ by carefully selecting $N_{\mathrm{val}}$. This approach could be advantageous for hardware setups that have reduced maximum voltage limits.

\begin{figure}
\includegraphics[width=\columnwidth]{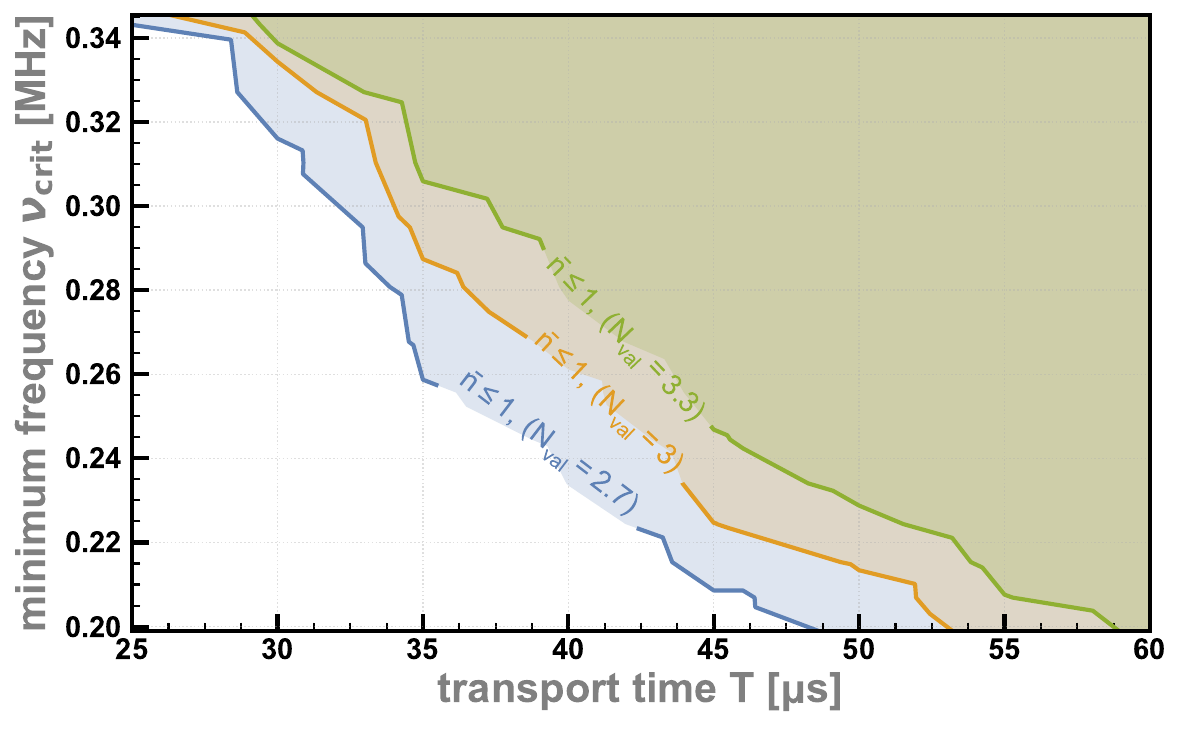}
\caption{Total transport time $T$ needed for \textit{near-adiabatic} two-ion splitting ($\bar{n}<1$), with respect to the minimum frequency $\nu_{\mathrm{crit}}$ and the slope parameter $N_{\mathrm{val}}$ of the transport profile. Analyzed for $N_{\mathrm{val}}=3.3$ (\textit{green}), $N_{\mathrm{val}}=3$ (\textit{yellow}) and $N_{\mathrm{val}}=2.7$ (\textit{blue}) in the context of an asymmetric frequency trajectory. The results in the case of merging are comparable, i.e., negligibly different.}
\label{fig:omegacomvsNval}
\end{figure}

This is further explored in Fig.~\ref{fig:omegacomvsNval}, where the asymmetric frequency trajectory is evaluated for different values of $\omega_{\mathrm{crit}}$ and $N_{\mathrm{val}}$. For example, the transition from the \textit{noisy} to the \textit{near-adiabatic} regime can be achieved within $\SI{33}{\mu s}$ either for $N_{\mathrm{val}}=2.7$ with $\omega_{\mathrm{crit}}\approx \SI{290}{kHz}$ or for $N_{\mathrm{val}}=3$ with $\omega_{\mathrm{crit}}\approx \SI{330}{kHz}$.

However, for reliably reaching the \textit{near-adiabatic} regime, the combination $\omega_{\mathrm{crit}}\approx \SI{345}{kHz}$ and $N_{\mathrm{val}}=3$ is consistently the most favorable choice (Fig.~\ref{fig:comparison_asym_heat}~b)). Taking into account that we aim to transport at time scales within the \textit{near-adiabatic} regime, we will consider this frequency interpolation (see Fig.~\ref{fig:frequency_trajectories} b)) and transport profile in the following. Additionally, this choice results in a frequency trajectory for which both the critical distance $d_{\mathrm{crit}}$ and the frequency interpolation can be constructed with our hardware specifications.

In the next section, we construct this exact frequency dependence and transport profile using our control voltages.

\subsection{ Voltage calculation} \label{sec: Voltage calculation}

In the following, we calculate the control voltages necessary to construct the frequency trajectory and transport profile, presented in the previous section. While it is common to split the transport path into equidistant spatial intervals, calculate voltage vectors for each position and then apply temporal shaping, here we calculate the transport voltages for a given set of time steps. Therefore, the trapping constraints are parametrized in the time domain. To do this, we first define the position constraints using the desired transport profile, $d(t)$ and plug it into Eq.~\eqref{eq:position merging}. Afterwards, we constrain the trapping frequency by inserting the desired frequency trajectory $\omega_{\rm{COM}}(t)$ into Eq.~\eqref{eq:curvature global merging}.

Consequently, this parametrization synchronizes the frequency trajectory and the transport profile, fulfilling the condition $d_{\mathrm{crit}}=d(t)\Big|_{t=T/4}$. And additionally $d_{\mathrm{in}}=d(t)\Big|_{t=T}$ as well as $d_{\mathrm{final}}=d(t)\Big|_{t=0}$ on the boundaries. These three equations define four vertices shown in Fig.~\ref{fig:transport profiles} (\textit{gray}).

\begin{figure}
\includegraphics[width=\columnwidth]{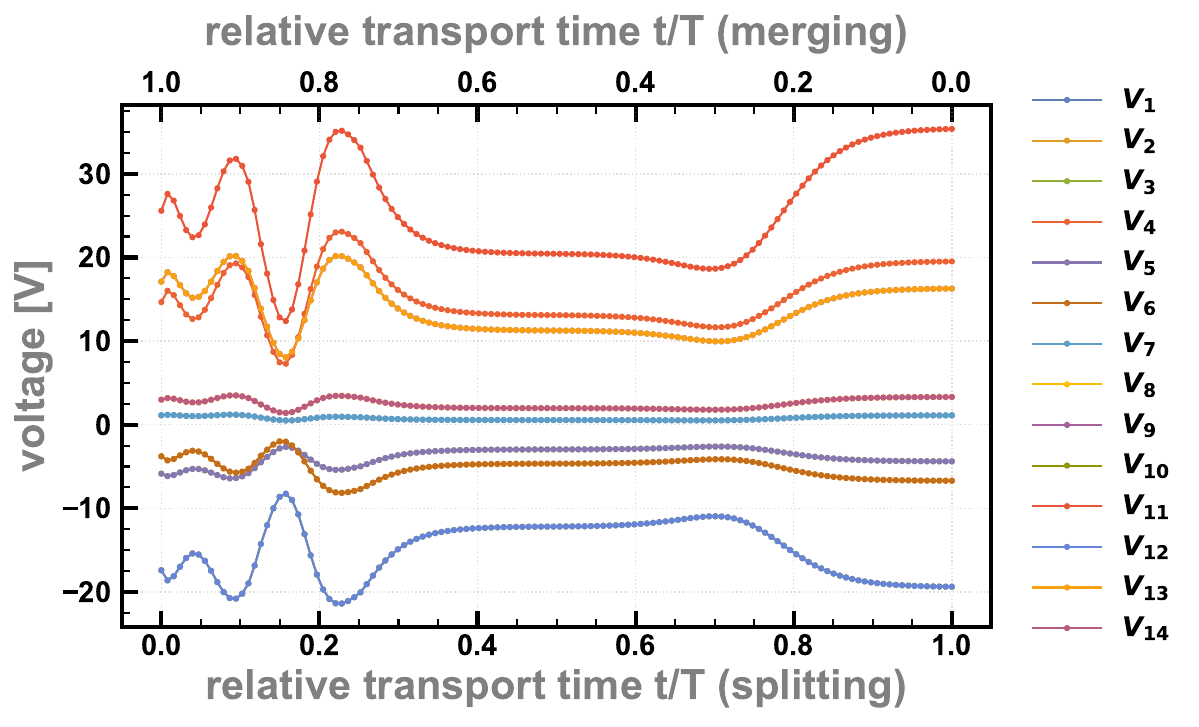}
\caption{Voltage waveforms for the 14 DC electrodes of our trap, parameterized in time for splitting and merging operations. \textit{Dots}: calculated positions, \textit{Lines}: interpolated values.}
\label{fig:voltage}
\end{figure}

An example of the resulting control voltages is presented in Fig.~\ref{fig:voltage}, where we calculate every $\frac{T}{200}\,\mu$s. 
Here, we use third-order polynomial interpolation to connect the calculation steps. 
There will therefore be deviations between the actual trajectory and the target transport profile $d(t)$. This can cause uncontrolled accelerations which add to the theoretically expected motional excitations presented in Fig.~\ref{fig:comparison_asym_heat}~a). 

\begin{figure}
\includegraphics[width=\columnwidth]{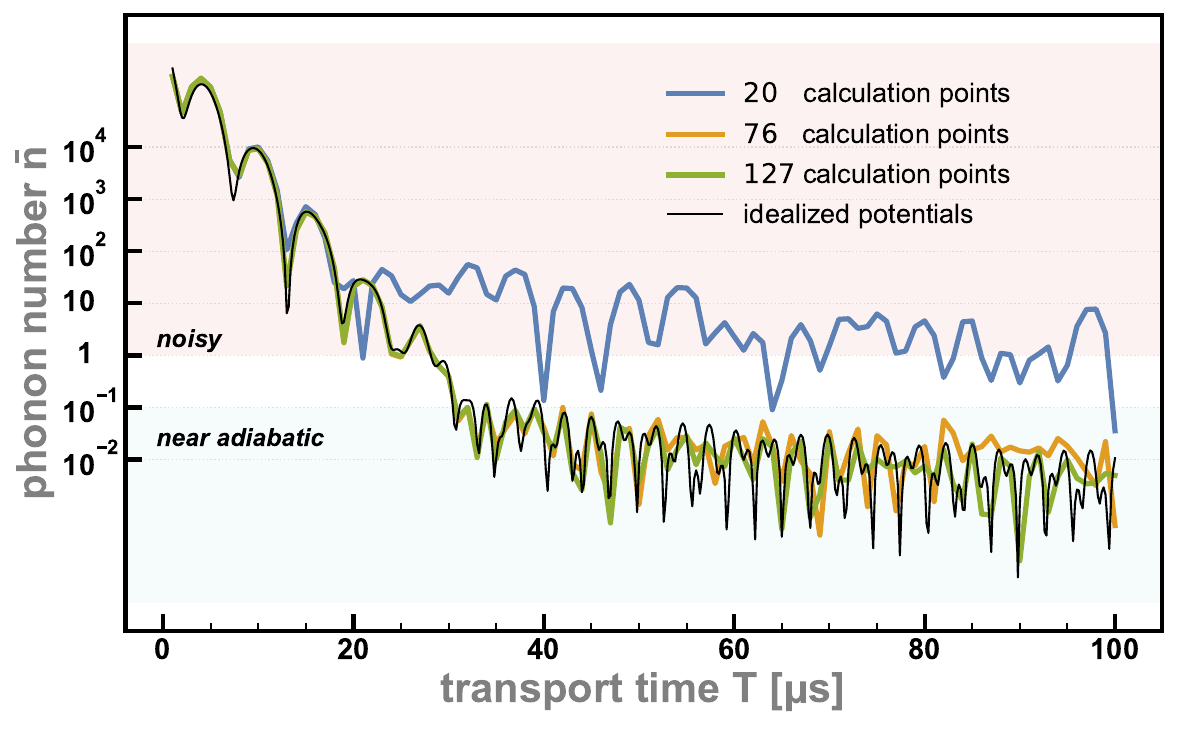}
\caption{Average phonon excitation vs. transport time in the context of two-ion splitting, considering the trapping potentials created by applying the voltage interpolation from Fig.~\ref{fig:voltage} to the electrode configuration from Fig.~\ref{fig: layout}. The asymmetric frequency trajectory $\nu_{\mathrm{com}}(t)$ from Fig.~\ref{fig:frequency_trajectories}~b) is used together with a $N_\text{val}$ parameter of 3 for the transport profile. The comparison to the graph in Fig.~\ref{fig:comparison_asym_heat}~b), which uses \textit{idealized potentials}, shows good agreement as more \textit{calculation points} are used. The remaining differences could originate from real-world anharmonicities~\cite{nuschkeAnalysisMotionalHeating2025a} and distortions in the target transport profile. The latter might arise from errors common in numerical differential equation solvers. The excitation in the case of merging is comparable, i.e., negligibly different.}
\label{fig:comparsion_interpol}
\end{figure}

To find the solutions, we solve the equation of motion as in Eq.~\eqref{eq:equations of motion_pot}:

\begin{equation}
\label{eq:equations of motion_pot}
    \begin{cases}
      \ddot{x}_1=-\frac{\partial\Phi_{\rm{st}}(x_1)}{\partial x_1}+\frac{\partial Cp(x_1,x_2)}{\partial x_1}\\ 
      \ddot{x}_2=-\frac{\partial\Phi_{\rm{st}}(x_2)}{\partial x_2}+\frac{\partial Cp(x_1,x_2)}{\partial x_2}\quad,
    \end{cases}
\end{equation}

where we use the electric field constructed in Sec.~\ref{sec: modelling and trap parameters}. Considering a specific electrode geometry allows us to paint a more realistic picture of the merging process, adding to the theoretical analysis presented in Sec.~\ref{sec:theoretical evaluation}. Besides that, this model allows us to include the effects of finite trap depths and analyze the influence of other effects, such as deviations resulting from the voltage interpolation. To test the sensitivity to the discretization of time, we compute the control voltages and solve the equation of motion in Eq.~\eqref{eq:equations of motion_pot} for different discretization levels and transport times. As shown in Fig.~\ref{fig:comparsion_interpol}, if the density of \textit{calculation points} decreases, $\bar{n}$ rises. This is a sensitive point for the physical implementation, since it establishes some requirements on the sampling rate of the hardware. Returning to the voltage waveforms, in Fig.~\ref{fig:voltage}, we observe a slope increase around $t<T/4$ (this refers to the timescale of the splitting process), which corresponds to $d<d_{\mathrm{crit}}$. These sudden changes are not captured correctly by the interpolation functions. It was observed that for 20 \textit{calculation points}, a maximum deviation of $< \SI{0.20}{\mu m}$ from the theoretical transport profile was registered. This highlights the sensitivity of the transport profiles.

Next, we present estimations of the sampling rates required to complete the operation in a given time $T$. Our hardware setup can provide a maximum sampling rate of $f_{\mathrm{s}}= \SI{2.55}{MS/s}$. To complete the merging process in $T\approx\SI{40}{\mu s}$, we can therefore use at most $T\times f_{\mathrm{s}}\approx 102$ interpolation points, using third-order interpolation in between samples. Effects of filters and other hardware components are not considered since these can be compensated for using reverse engineering~\cite{vonboehnSpeedingAdiabaticIon2025,qiOptimizingIonShuttlingOperations2021,kriegerSpeedingIonTransport2026,contaToolchainShuttlingTrappedion2026}.

\section{Two ion swapping} \label{sec:swap}

\subsection{Geometrical considerations for two-ion swapping \label{sec: Geometrical considerations for two ion swapping}}
Swapping operations are achieved by rotating the principal axes of a 3-dimensional harmonic potential by $180^\circ$. While this technique can be applied to setups with multiple ions, which may be of different species~\cite{delaneyScalableMultispeciesIon2024}, here we focus on the simpler case of two equal-mass ions. This simple case has been demonstrated and well characterized in different setups~\cite{kaufmannScalableQuantumProcessor2017,splattDeterministicReordering40Ca2009a,vanmourikCoherentRotationsQubits2020a}.

In the following, we analyze the resulting requirements on the DC-electrode geometry. The rotation of the principal axes can be done in a plane that includes the $x$-axis. Depending on the particular geometry, constant secular frequencies can be maintained, if desired. This is feasible since the total curvature available at the point of calculation remains constant during the rotation~\cite{blakestad_transport_2010}. Consider a rotation in the $x$-$y$ plane. Hence half way into the process the crystal will be aligned with the $y$-axis. To maintain a constant axial frequency, a large anti-confining static potential along $y$ will be required to compensate for the strong radial pseudopotential confinement. Whether or not this is possible strongly depends on the DC-electrode geometry. Since the ion-to-electrode distance is fixed, the main influence is from the electrode widths. These electrodes need to be wide enough to avoid exceeding the voltage range limit without compromising the performance of merging and splitting operations or exceeding the constrained gate zone size.

\subsection{Constraints and waveform calculations}
One might be tempted to apply the procedure, described in Sec.~\ref{sec: system constraints}, where the two ion positions are individually constrained. However, as we are using an asymmetric RF electrode, strong potential asymmetries are present at the trapping position~\cite{homeNormalModesTrapped2011}. 
This complicates the use of two-point constraints, particularly when the ions are aligned with the global y-axis, since constraining in two equidistant points implies obtaining a symmetric total potential. This is however not feasible within our voltage limit. Hence here we use single-point constraints at the trap center and rotate the principal axes by an angle $\theta$. Nevertheless, using single-point constraints results in an asymmetric total potential where the ions are not equidistant from the trap center with corresponding changes in the mode frequencies and principal axes~\cite{homeNormalModesTrapped2011}. In a first attempt, it was observed that rotating while maintaining a constant base axial curvature at the trap center resulted in the ions approaching anti-confining positions. This occurs due to the high pseudopotential asymmetry along the $y$-axis. To gain more control over the ion positions, we symmetrically vary the local axial curvature at the constrained position as the rotation occurs. We start with a base axial curvature and increase its value until the point where the two ions are align with the $y$-axis, $\theta=90^\circ$. The other half of the rotation from $\theta=90^\circ$ to $\theta=180^\circ$ is mirror symmetric with respect to the first half around $\theta=90^\circ$. This transient increase in curvature brings the ions closer together, away from anti-confining positions. In more detail, to influence the mode frequencies, we apply the following constraints to the total potential:

\begin{equation}
\label{eq:rotation}
 R^{-1}(\theta)\cdot H(\phi_{\mathrm{tot}}(r_c))\cdot R(\theta)\doteq D
\end{equation}

\begin{figure}
    \centering
    \includegraphics[width=\columnwidth]{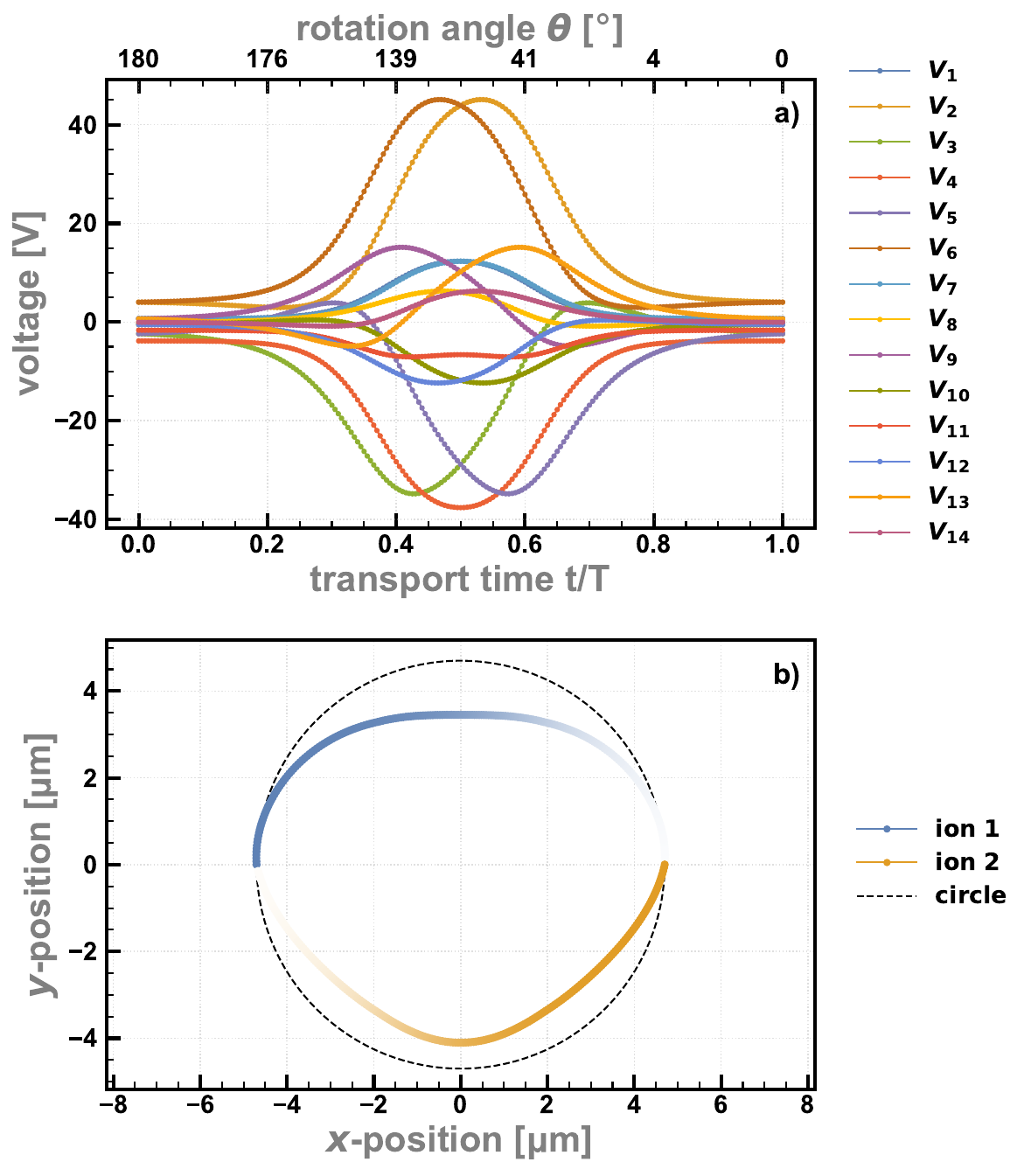}
    \caption{\textit{a}) Calculated transport waveforms for the 14 electrodes of our trap to implement a two-ion swapping operation, parameterized in time with 200 support points.
    \textit{b}) Equilibrium positions $r_1$ and $r_2$ of the two ions during the rotation. A circular trajectory is shown for comparison (\textit{black, dashed}). This corresponds to a rotation at a constant $\omega_\text{COM}=1\, $MHz.}
    \label{fig:swapping_trajectory}
\end{figure}

We transform the total potential Hessian $H(\phi_{\mathrm{tot}}(r_c))$ with a 3x3 matrix, $R(\theta)$, which is parametrized to rotate around the $z$-axis. Here, $D$ is a diagonal matrix containing the eigen-curvatures\footnote{Since this is a position where none of the ions are, a conversion to frequency is not appropriate.}, which allow us to influence the crystal-mode frequencies and approximate the equilibrium positions. Furthermore, we constrain the trap minimum, $r_c$, with $\nabla\phi_{\mathrm{tot}(r_c)}\doteq0$. Afterwards, we solve the system of equations and obtain the voltage waveforms. To obtain time-dependent voltages, we parametrize $\theta$ by a single hyperbolic tangent\footnote{Similarly to Sec.~\ref{sec:merging and splitting}, different $N_{\mathrm{val}}$ parameters were tested. Yet the same trend was observed: larger $N_{\mathrm{val}}$ parameters result in smaller excitations, but after larger transport times. $N_{\mathrm{val}}=3$ is again a reasonable compromise. A finer exploration around this value was not presented since the axial frequency is considerably higher than during the merging operation.} with $N_{\mathrm{val}}=3$. The resulting time dependent waveforms can be seen in Fig.~\ref{fig:swapping_trajectory}~a), where the highest slopes are located around $\frac{1}{2}T$.

After calculating the voltage waveforms, we find the global coordinates of the equilibrium positions, $r_1$ and $r_2$, where their $x$ and $y$ components are shown in Fig.~\ref{fig:swapping_trajectory}~b). As expected, the Euclidean distance between the ions changes while rotating. The combination of the varying central trapping curvature and the pseudopotential asymmetry results in the deviations from a circular trajectory.

\begin{figure}
    \centering
    \includegraphics[width=\columnwidth]{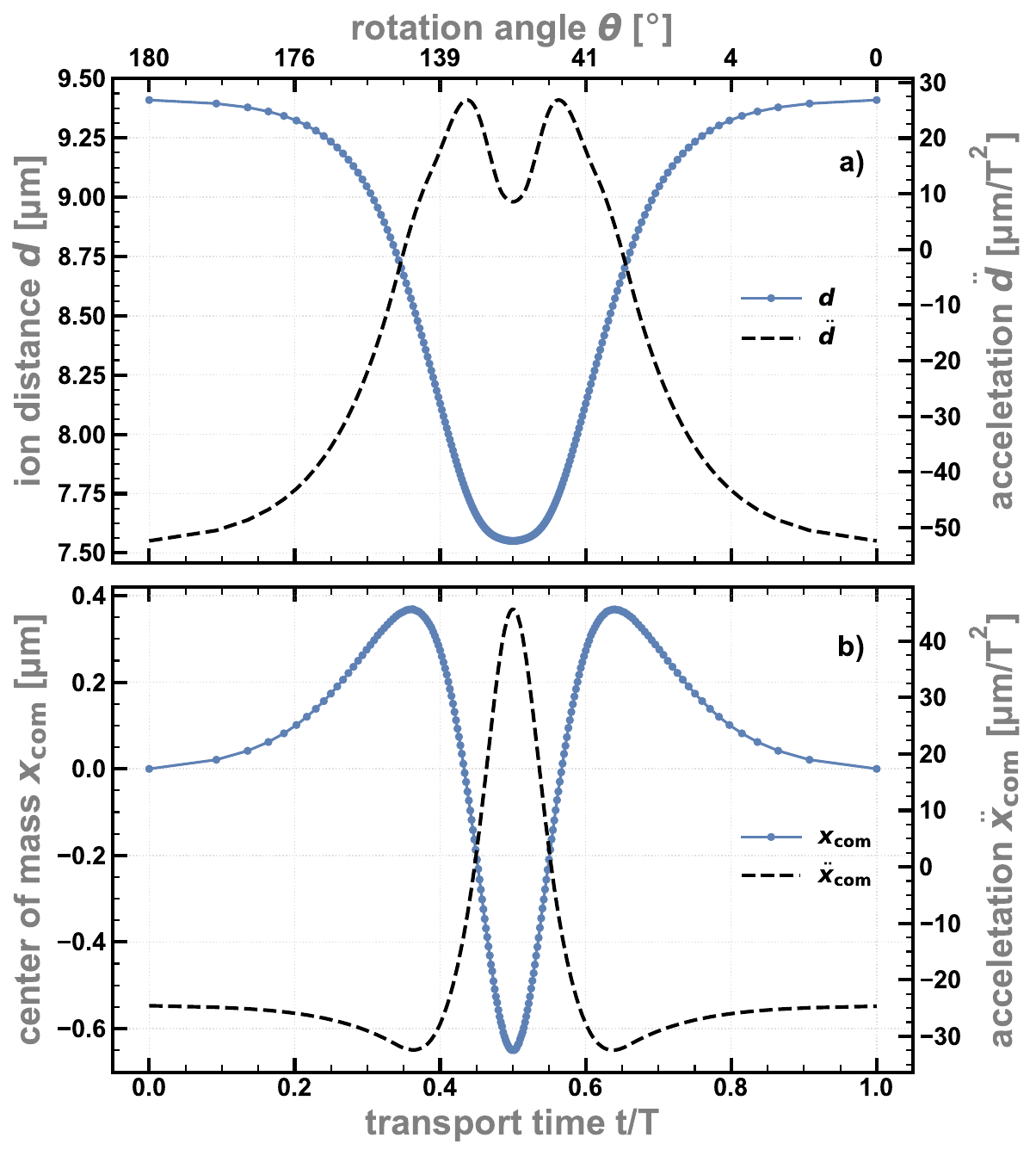}
    \caption{Equilibrium positions in a rotating frame during the two-ion swapping. \textit{a}) Distance between ions. \textit{b}) Center position relative to the axis of rotation. In a perfectly executed swapping operation (see Fig.~\ref{fig:swapping_trajectory} \textit{circle}), based on purely harmonic rotating potentials, we should have $\partial d/ \partial t=x_\text{COM}(t)=0$.}
    \label{fig:trajectory_relative}
\end{figure}

\subsection{Equations of motion and energy transfer}

\begin{figure}
\includegraphics[width=\columnwidth]{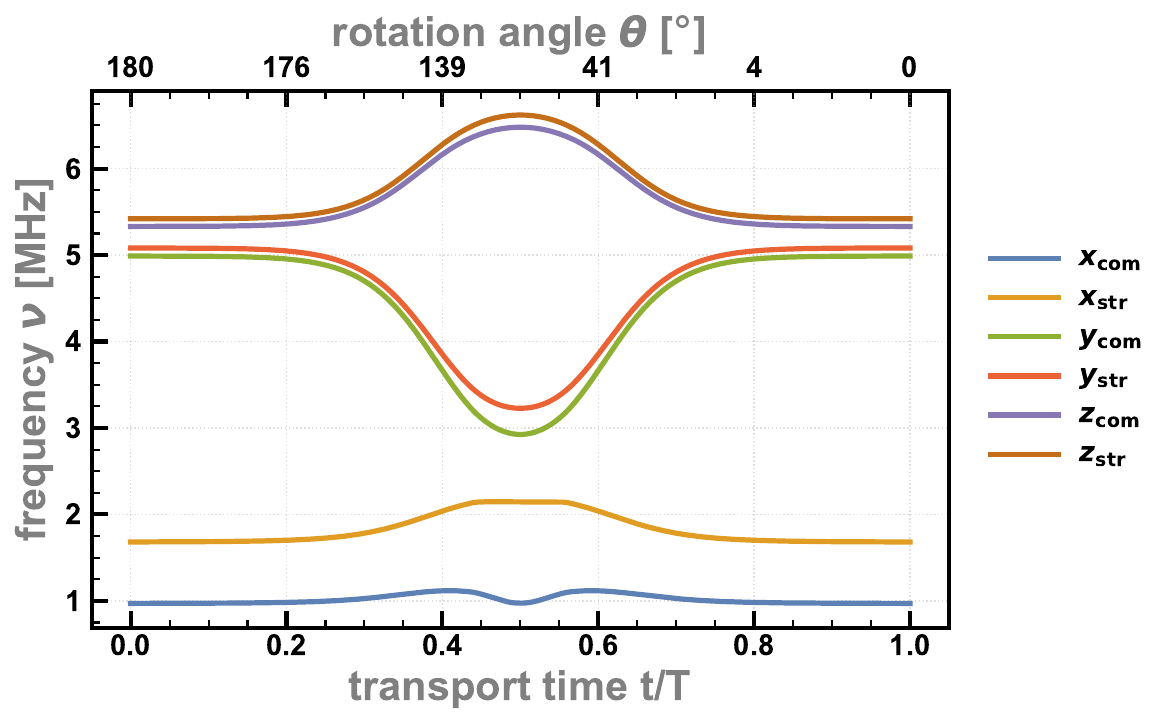}
\caption{IP and OOP mode frequencies of the two-ion crystal during the swapping operation. Changes in the axial COM mode are small compared to those in the STR mode frequency. Changes in distance are compensated by the changing pseudopotential hexapole term and center curvature.}
\label{fig: secular frequencies}
\end{figure}

Since we have limited control over the equilibrium positions and secular frequencies, we next analyze the magnitudes of the IP (in-phase) and OOP (out-of-phase) modes. This allows us to check that we do not have degeneracy between the radial and axial secular modes, which could lead to population exchange \cite{delaneyScalableMultispeciesIon2024}. From the expected mode frequencies shown in  Fig.~\ref{fig: secular frequencies}, we make the following observations:

\begin{itemize}
  \item The axial STR frequency is more sensitive to changes in the equilibrium positions compared to the COM frequency. This is due to the strong dependency of the STR mode on the equilibrium distance $\propto 1/d^3$, while the COM mode depends on the potential curvature at $r_1$ and $r_2$. To offer a different perspective, we move to a relative frame along the axial direction, where we consider a 1-dimensional potential. For this we use the potential in Eq.~\eqref{eq:quartic potential} with $\beta=0$ and $\gamma \neq 0$. Analogously to Sec.~\ref{sec: system constraints} at the equilibrium distance the curvature is given by the relations between, $\alpha$ and  $\gamma$. This curvature results, in this case, in values near $\SI{1}{MHz}$. These relations are analyzed in more detail in~\cite{homeElectrodeConfigurationsFast2006a}.
  
  \item Similarly, Fig.~\ref{fig: secular frequencies} shows the radial modes, where one can observe changes in the IP and OOP frequencies. The changes in the IP frequency can be attributed to the unconstrained condition of the radial modes. Moreover the changes in the OOP mode frequencies are mediated by the distance in between the ions and the IP mode frequencies as expected from Eq.~\eqref{eq: STRvsCOM}.
\end{itemize}

Since we have a minimum difference of approximately $\SI{1}{MHz}$ between the axial and radial modes, we do not expect population exchange and continue with the heating analysis. Therefore we proceed to calculate $\bar{n}$, using the equations of motion. Swapping operations require temporarily moving away from the RF-null line (pseudopotential minima) to positions where the motion is also driven by RF fields (excess micromotion). The influence of excess micromotion can be studied by including the RF potential in the equations of motion (see Eqs.~\eqref{eq:app_expansion} and~\eqref{eq:app_Ht}); however, the precise effects still need to be studied and are therefore outside the scope of this work. For our purposes of establishing a principle for considering phonon excitations at the compiler level, we will use the pseudopotential approximation in the following.

\begin{equation}
\label{eq:rot equations of motion pot}
    \begin{cases}
      \ddot{x}_1=\frac{-\partial\Phi_{\mathrm{tot}}(x_1,y_1)}{\partial x_1}+\frac{\partial Cp(x_1,x_2,y_1,y_2)}{\partial x_1} 
      \\ 
      \ddot{x}_2=\frac{-\partial\Phi_{\mathrm{tot}}(x_2,y_2)}{\partial x_2}+\frac{\partial Cp(x_1,x_2,y_1,y_2)}{\partial x_2}
      \\
      \ddot{y}_1=\frac{-\partial\Phi_{\mathrm{tot}}(x_1,y_1)}{\partial y_1}+\frac{\partial Cp(x_1,x_2,y_1,y_2)}{\partial y_1}
      \\ 
      \ddot{y}_2=\frac{-\partial\Phi_{\mathrm{tot}}(x_2,y_2)}{\partial y_2}+\frac{\partial Cp(x_1,x_2,y_1,y_2)}{\partial y_2}
    \end{cases}
\end{equation}

To capture all forces influencing the ions trajectories we solve the equations of motion using global coordinates (see. Eq.~\eqref{eq:rot equations of motion pot}). Similar to Sec.~\ref{sec:merging and splitting}, we obtain the total energy transfer from Eq.~\eqref{eq:motional excitation}.

\begin{figure}
    \centering
    \includegraphics[width=\columnwidth]{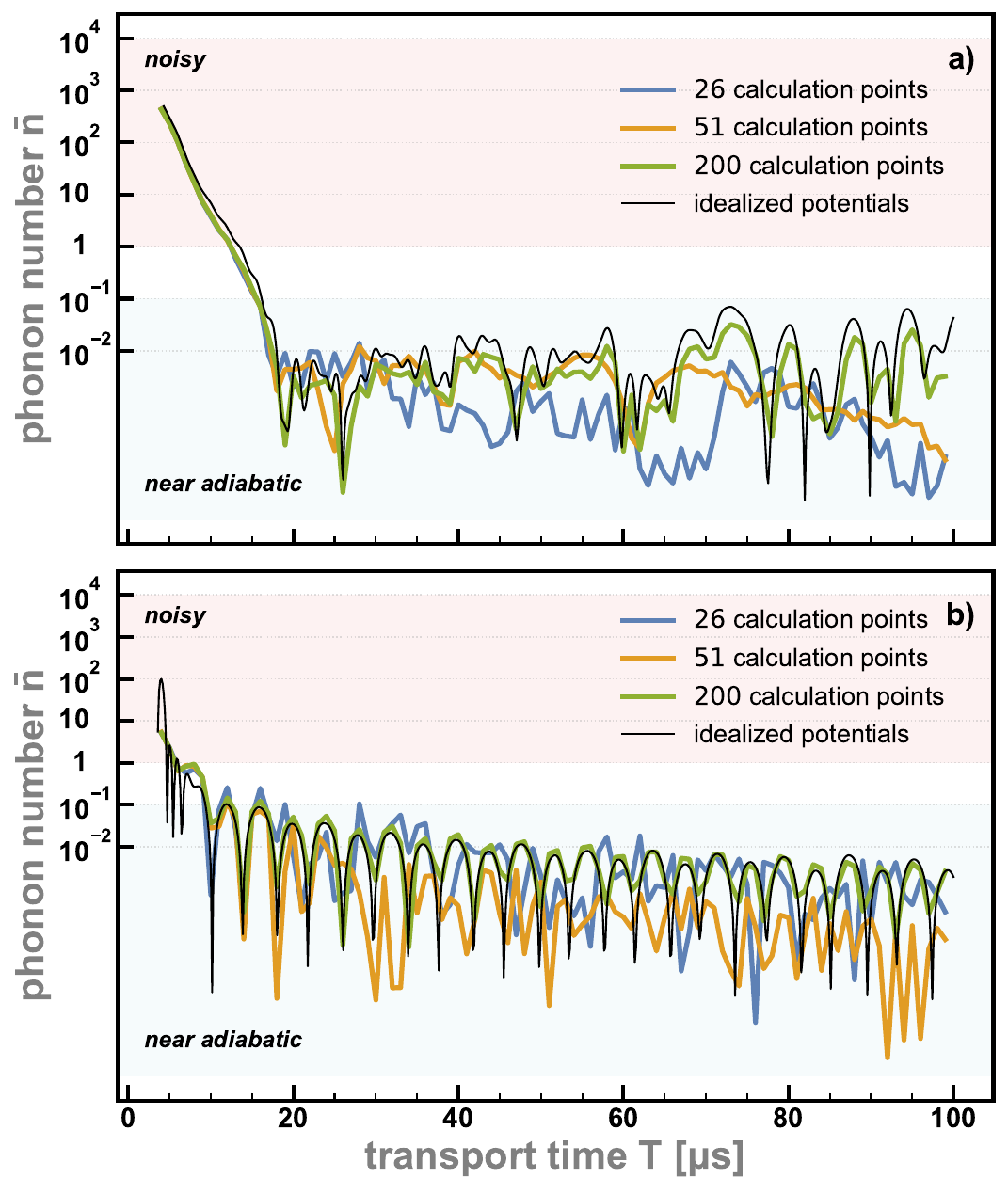}
    \caption{Average phonon number of the axial modes vs. transport time in the context of two-ion swapping, considering a real potentials created by applying the calculated control voltages to the electrode configuration in Fig.~\ref{fig: layout}. Shown is the response to different numbers of \textit{calculation points} (i.e., positions where the voltages (see Fig.~\ref{fig:swapping_trajectory}) have been evaluated), as well as control curves based on \textit{idealized potentials}. \textit{a}) Excitation of the COM mode, $\bar{n}_{com}$. \textit{b}) Excitation of the STR mode, $\bar{n}_{str}$. A transport profile with $N_\text{val}=3$ is used. The control curves show good agreement as more \textit{calculation points} are used. Remaining differences in the phonon excitation might be explained by small real-world anharmonicities~\cite{nuschkeAnalysisMotionalHeating2025a} induced by the electrode geometry. Other possible reasons are distortions in the target transport profile caused by interpolations between \textit{calculation points}.}
    \label{fig:heating_rot}
\end{figure}

In Fig.~\ref{fig:heating_rot} we observe excitations in both normal axial modes, which originate from different sources. While changes in the relative distance of the two ions ($d=u_1-u_2$) induce phonons in the STR mode, variations in their center positions ($x_{\mathrm{com}}=(u_1+u_2)/2$) excite the COM mode (see Fig.~\ref{fig:trajectory_relative}, \textit{solid}). Moreover, it can be noticed that the excitation of the COM mode is the dominant contribution to the phonon number. Though the magnitudes of the accelerations (see Fig.~\ref{fig:trajectory_relative}, \textit{dashed}) are comparable, the profiles look vastly different. Since the motional heating originates from the resonance condition between the trapping frequency and the Fourier components of the transport acceleration profile~\cite{reichleTransportDynamicsSingle2006}, different excitations in the cases of $d$ and $x_{\mathrm{com}}$ are not unexpected. Further explanation is provided by the difference in the trapping frequencies, as larger frequencies are generally less susceptible to being resonant with the Fourier components of the transport acceleration profile~\cite{nuschkeAnalysisMotionalHeating2025a}.

Furthermore, to identify the individual influence of inertial forces acting on the ions while rotating, we analyze the motion in a rotating frame. We characterized the individual effects, by sequentially incorporating the Coriolis, centrifugal, and Euler forces into the simulation. It was observed that their contributions are only relevant for $T<1\,\mu$s. Hence, for the time scales of interest, changes in the equilibrium positions due to the trapping potential are dominant.

Large excitations of both modes are partly not registered because of the relative smoothness of the waveforms. These let the interpolation function accurately capture the voltage slopes. Conversely to Section~\ref{sec:merging and splitting}, we observe less sensitivity to the interpolation points.

Finally, it was observed that the swapping process results in motional excitation on the radial modes with a maximum of $\bar{n}_y\approx0.01$.

\section{Operation compiler} \label{sec:Operation compiler}

In the following, we characterize the operation of the chip, considering the above discussed different ion manipulations possible on our platform. To do this, we first present a methodology to quantify the motional excitation induced by ion reordering. This methodology is based on decomposing the full reordering routine into a sequence of primitive operations, as each of these primitive operations can be associated with a characteristic heating (see Sec.~\ref{sec:single ion}). Afterwards, using the configuration shown in Fig.~\ref{fig: layout}, we quantify different reordering operations in terms of the phonon excitations and operation times.

\subsection{Classification of Heating} \label{sec:classheat}

As discussed in Sec.~\ref{sec:Chip Design}, our primitive operations include transport, merging, splitting, and swapping. As seen in Sec.~\ref{sec:merging and splitting} and Sec.~\ref{sec:swap}, we found the characteristic heating by simulating the motion of each operation with a given set of trapping parameters. However, to find the total phonon excitation after an entire routine is completed, we do not wish to simulate all the successive operations from start to end. This would become computationally expensive as the number of operations grows. Here, we rather reconstruct the total motional excitation from the individual phonon excitations of the four primitive operations. If the primitive operations are pre-calculated for the desired times and distances, this approach simplifies the computation process, as all information needed is the resulting motional excitation induced by each primitive operation. The reconstruction of the total motional excitation is then obtained by using the corresponding contributions.

\begin{figure}
\includegraphics[width=\columnwidth]{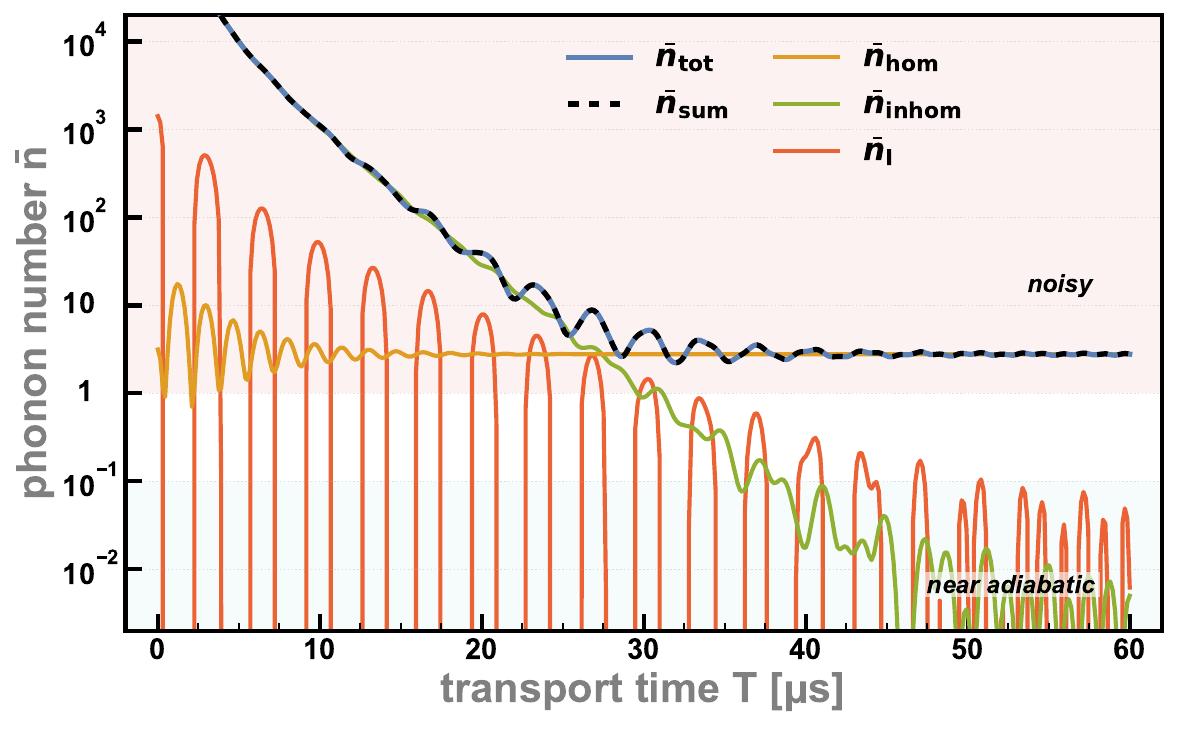}
\caption{Phonon excitation during two-ion merging with initial excitations, decomposed into the contributions of the harmonic solution $\bar{n}_{\mathrm{hom}}$, the anharmonic solution $\bar{n}_{\mathrm{inhom}}$, and the interference term $\bar{n}_{I}$. The sum $\bar{n}_{\mathrm{sum}}$ of these components corresponds precisely to the heating resulting from the complete solution. A hyperbolic tangent with $N_{\mathrm{val}}=3$ is selected for the transport.}
\label{fig:different_contributions_to_n}
\end{figure}

However, it must also be taken into account that the excitation resulting from the operations preceding the $i$-th step, $\bar{n}_{i-1,\ \text{tot}}$ (i.e. the total excitation after the first $i-1$ operations), generally does not add linearly to the excitation generated solely by the $i$-th operation\footnote{This is what we will later refer to as the inhomogeneous solution of the $i$-th operation.}, as an interference term $\bar{n}_I$ additionally arises (Fig. \ref{fig:different_contributions_to_n} shows the separate contributions to the phonon number). This can be quantified by decomposing the trajectory of the $i$-th operation into its homogeneous and inhomogeneous parts ($x_{i,\ \text{tot}}=x_{i,\ \text{hom}}+x_{i,\ \text{inhom}}$). From this, we obtain the following expression. All quantities in the derivation refer to the same step $i$ of the cascade. The index is therefore suppressed and reinstated in Eq.~\eqref{eq:recursive2}.
\begin{equation}\label{5.12}
    \begin{aligned}
    \bar{n}_{\text{tot}} &= \frac{1}{\hbar \omega_0} \left[ 
    \frac{1}{2} m (\dot{x}_{\text{hom}} + \dot{x}_{\text{inh}})^2 + \frac{1}{2} m \omega_0^2 (x_{\text{hom}} + x_{\text{inh}})^2\right] \\
    &= \bar{n}_{\text{hom}}
    + \bar{n}_{\text{inh}}+ \underbrace{\frac{1}{\hbar \omega_0} \left( m \dot{x}_{\text{hom}} \dot{x}_{\text{inh}} + m \omega_0^2 x_{\text{hom}} x_{\text{inh}} \right)}_{\bar{n}_{I}}
    \end{aligned}
\end{equation}
Given that the potentials in our trap are harmonic to a very good approximation, the solutions at the end of a transport can be expressed in the form\footnote{This is a parametrization of the final phase-space coordinates, rather than implying a subsequent hold time: the pair $(x_k,\dot{x}_k)$ at the end of the transport determines $A_k$ and $\varphi_k$. The oscillation is in fact already visible before the nominal end of the operation, as the transport profiles used here gradually approach their final position.}:
\begin{equation}
    x_k(t) = A_k \cos(\omega t + \varphi_k),\
    \dot{x}_k(t) = -\omega_0 A_k \sin(\omega t + \varphi_k),
\end{equation}
with $k \in \{\text{hom},\, \text{inh}\}$. Therefore, using the trigonometric identity $\cos\alpha\cos\beta + \sin\alpha\sin\beta = \cos(\alpha-\beta)$, the interpolation term can be written as:
\begin{equation}\label{eq:interf}
    \begin{aligned}
    \bar{n}_{I}&= \frac{m \omega}{\hbar} A_{\text{hom}} A_{\text{inh}}
    \cos\!\big(\varphi_{\text{hom}} - \varphi_{\text{inh}}\big)\\
    &= 2 \sqrt{\bar{n}_{\text{hom}}\, \bar{n}_{\text{inh}}} \, \cos\vartheta,\quad\vartheta=\varphi_{\text{hom}} - \varphi_{\text{inh}},
    \end{aligned}
\end{equation}
leading to:
\begin{equation}
\label{eq:recursive2}
\begin{aligned}
    \bar{n}_{i,\ \text{tot}} &= \bar{n}_{i,\ \text{hom}}(\bar{n}_{i-1,\ \text{tot}},\vartheta_{i-1}) + \bar{n}_{i,\ \text{inhom}} \\
    &\quad + \underbrace{2\sqrt{\bar{n}_{i,\ \text{hom}}(\bar{n}_{i-1,\ \text{tot}},\vartheta_{i-1}) \cdot \bar{n}_{i,\ \text{inhom}}} \cdot \cos(\vartheta_i)}_{\bar{n}_{i,\ I}(\bar{n}_{i,\ \text{hom}},\ \bar{n}_{i,\ \text{inhom}})}
\end{aligned}
\end{equation}
The recursive formulation in Eq.~\eqref{eq:recursive2} solely depends on the initial phonon excitation, the excitation induced by the current operation, and the respective phase relations between the homogeneous and inhomogeneous solutions \footnote{i.e., the phases at the beginning of the $i$-th operation, labeled here as $\vartheta_{i-1}$. Since a transport operation usually does not address all ions at the same time, there are actually five primitive operations, whereby the fifth does nothing other than cause the corresponding ion to wait and account for the phase evolution while other ions are manipulated.}. All but one of these quantities can be extracted from the four individual heating analyses of the primitive operations.

The only exception is the homogeneous contribution $\bar{n}_{i,\ \text{hom}}(\bar{n}_{i-1,\ \text{tot}},\vartheta_{i-1})$, which must be evaluated based on the initial conditions — and therefore has to be recalculated at each intermediate step of the full routine. However, this can be circumvented by constructing a fundamental matrix $\Theta_T(t)$ for a basis of initial conditions. This then obeys $x_{\mathrm{hom},\ T}(t)=\Theta_T(t)\cdot x_0$, with $x_0$ denoting the known initial conditions. If needed, interpolating $\Theta_T(t)$ over the total transport time $T$ yields a matrix function $\Theta(t,T)$, allowing one to manipulate the total transport time, in addition to the initial conditions, without resolving the differential equations. This fundamental-matrix ansatz cannot be applied to the total solution $x_{i,\ \text{tot}}$, since the latter belongs to an affine rather than a linear solution space and therefore undergoes affine, instead of linear, propagation.

\begin{figure}
\includegraphics[width=\columnwidth]{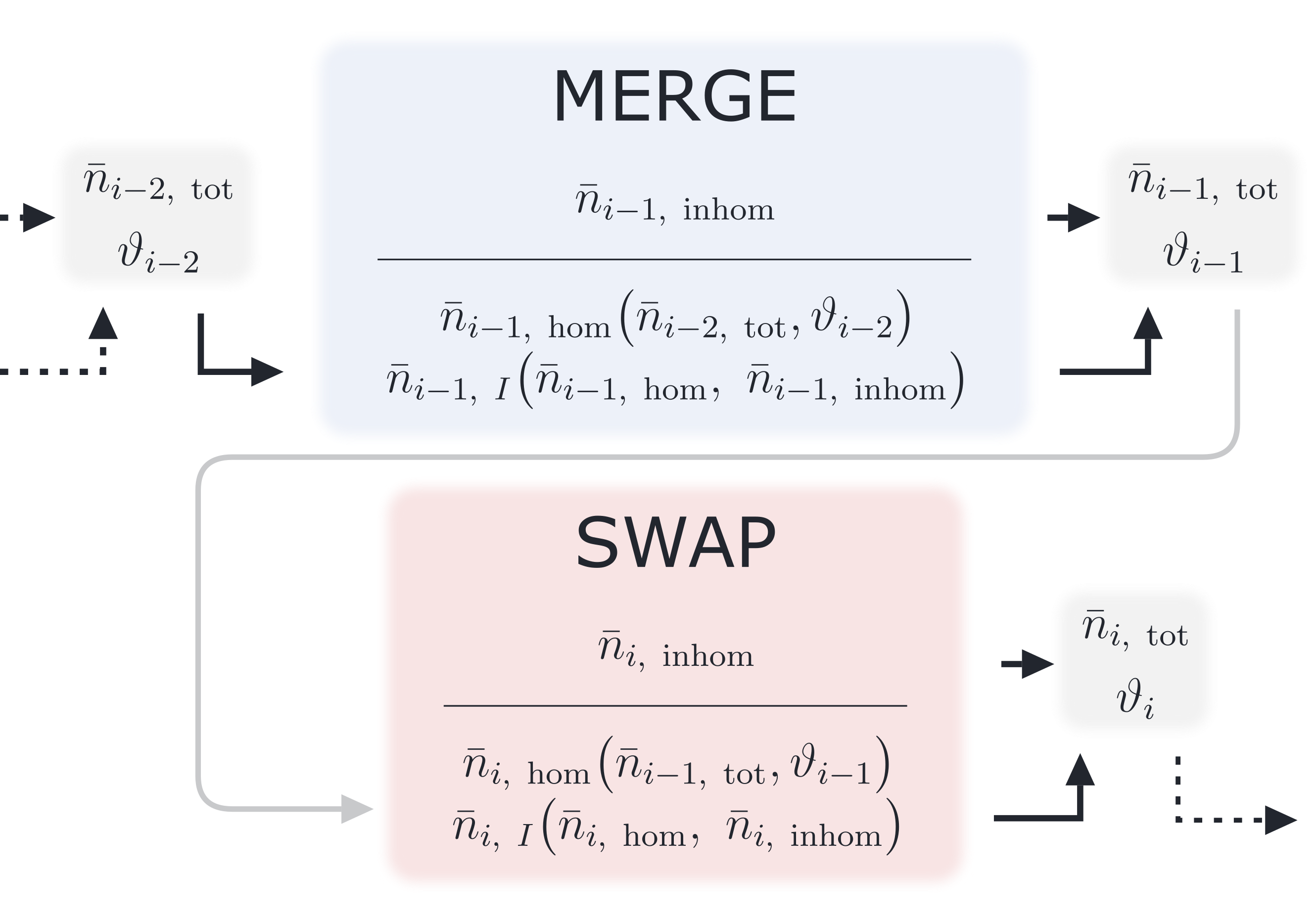}
\caption{Breakdown of the term structure within a hypothetical shuttling sequence. The inhomogeneous solutions depend exclusively on the primitive operations evaluated in advance. Only the homogeneous solutions, and thus the interference term, are determined by the preceding sequence of operations.}
\label{fig:overview_terms}
\end{figure}

Therefore, the heating according to the reordering routine might be decomposed into single contributions, which can be called via predefined functions. Evaluated in the order of the routine, this gives the total amount of heating accumulated due to the reordering. An overview of the calculation scheme for successive operations is given in Fig.~\ref{fig:overview_terms}. Note that we are working within the general description of the pseudopotential approximation. However, our model is not specifically limited to this approximation, this is clarified mathematically in Appendix~\ref{app:fundamental_matrix}.

What we have implicitly assumed at this point is that the phonon excitation caused by the transport is the same when starting from the ground state as when starting from an arbitrarily excited state. This assumption is indeed valid, at least with respect to the excitation induced by acceleration, which we have limited ourselves to here (see~\ref{sec:appendix_initial_fock}).

The recursion in Eq.~\eqref{eq:recursive2} can also be given in compact closed form. To this end we may collect the secular motion and phase of each contribution into a complex amplitude
$\mathcal{A}=\sqrt{\frac{m\omega}{2\hbar}}\left(x+\frac{\mathbf{i}\,\dot x}{\omega}\right)=\sqrt{\bar n}\,e^{\mathbf{i}\,\varphi}$, so that $\mathcal{A}_{\mathrm{tot}}=\mathcal{A}_{\mathrm{hom}}+\mathcal{A}_{\mathrm{inhom}}$ and $\bar n_{\mathrm{tot}}=\big|\mathcal{A}_{\mathrm{tot}}\big|^{2}$. With
$\Theta_{i}$ understood to act on the amplitude, one step reads $\mathcal{A}_{i,\mathrm{tot}}=\Theta_{i}\big[\mathcal{A}_{i-1,\mathrm{tot}}\big]
+\mathcal{A}_{i,\mathrm{inhom}}$, and repeated insertion with $\mathcal{A}_{0}=0$ gives
\begin{equation}
\label{eq:cascade_closed}
\begin{aligned}
\bar n_{N,\mathrm{tot}}&\;=\;\big|\mathcal{A}_{N,\mathrm{tot}}\big|^{2}\\
  &\;=\;\Bigg|\sum_{i=1}^{N}
    \big(\Theta_{N}\!\circ\!\Theta_{N-1}\!\circ\!\cdots\!\circ\!\Theta_{i+1}\big)
    \big[\mathcal{A}_{i,\mathrm{inhom}}\big]\Bigg|^2.
\end{aligned}
\end{equation}

\subsection{Compiler Network and Operation} \label{sec:network}
Knowing how to account for the induced heating by applying a given manipulation sequence to a certain ion configuration, we now need to find the ideal sequence in terms of operation time and induced phonon excitation.

Finding the complete network (i.e. all possible manipulation sequences, see Fig.~\ref{fig:network}) of ion manipulations for a given reordering task would allow the formulation of a linear optimization problem, which could be used, for example, to determine the minimum number of transport reordering operations; this solution would constitute a global minimum. However, considering the number of nodes, this is computationally infeasible. Similarly, using search algorithms with no heuristics would amount to a time complexity of $\mathcal{O}(V+E)$, where $V$ is the number of nodes and $E$ is the number of edges.

Therefore, here we apply a weighted branch search with heuristics.
~This type of search typically converges towards a local minimum.

Consider an instance where we start with the configuration depicted in Fig.~\ref{fig: layout} and wish to bring ion 1 and 8 to the gate zone. This is a particularly expensive operation, since the effective distance between the ions is rather large. An unweighted search, which minimizes only the number of operations, finds a local minimum path with 54 operations (6 swapping operations and 13 merging / splitting operations).

However, as already pointed out in the introduction, the sheer number of operations is not a comprehensive measure of the performance of a particular design. Hence one has to perform a weighted search and consider the performance of the individual operations. 

First, we will quantify the performance by performing each operation in a \textit{noisy} regime (here, $\bar{n} \approx 1$). After targeting a given transport time, we make use of the inhomogeneous solutions presented in Sec.~\ref{sec:merging and splitting} and Sec.~\ref{sec:swap} together with the methodology presented in Sec.~\ref{sec:classheat} to calculate the motional excitation of all ions after each elementary operation.

Table~\ref{table:values_compiler} shows the number of calculation points required by each elementary operation to achieve the \textit{near-adiabatic} and \textit{noisy} regimes, when using the maximum sampling frequency that our hardware can provide.

\begin{table}
  \caption{Summary of values in transport time and phonon number for two-ion shuttling operations. For single-ion transport, we use $12,\mu$s to ensure excitations below 1 phonon and $14,\mu$s for excitations below 0.1 phonons ($N_{\mathrm{val}}=5$).}
  \label{table:values_compiler}
  \begin{tblr}{
		width=\textwidth,
		colspec={Q[1.5cm,j,t] Q[0.9cm,c,t] Q[0.9cm,c,t] Q[2cm,c,t] Q[0.9cm,c,t]}
  }
\hline\hline
Operation   & $\bar{n}$      & $T$        & {Interpolation \newline points} & $N_{\mathrm{val}}$ \\ \hline
Merge/Split & 1              & $30\,\mu$s & 76                              & 3\\  
Swap        & 1              & $18\,\mu$s & 45                              & 3\\  
Merge/Split & 0.1            & $40\,\mu$s & 100                             & 3\\  
Swap        & 0.1            & $20\,\mu$s & 50                              & 3\\
\hline\hline
  \end{tblr}
\end{table}

\begin{figure}
    \centering
    \includegraphics[width=\columnwidth]{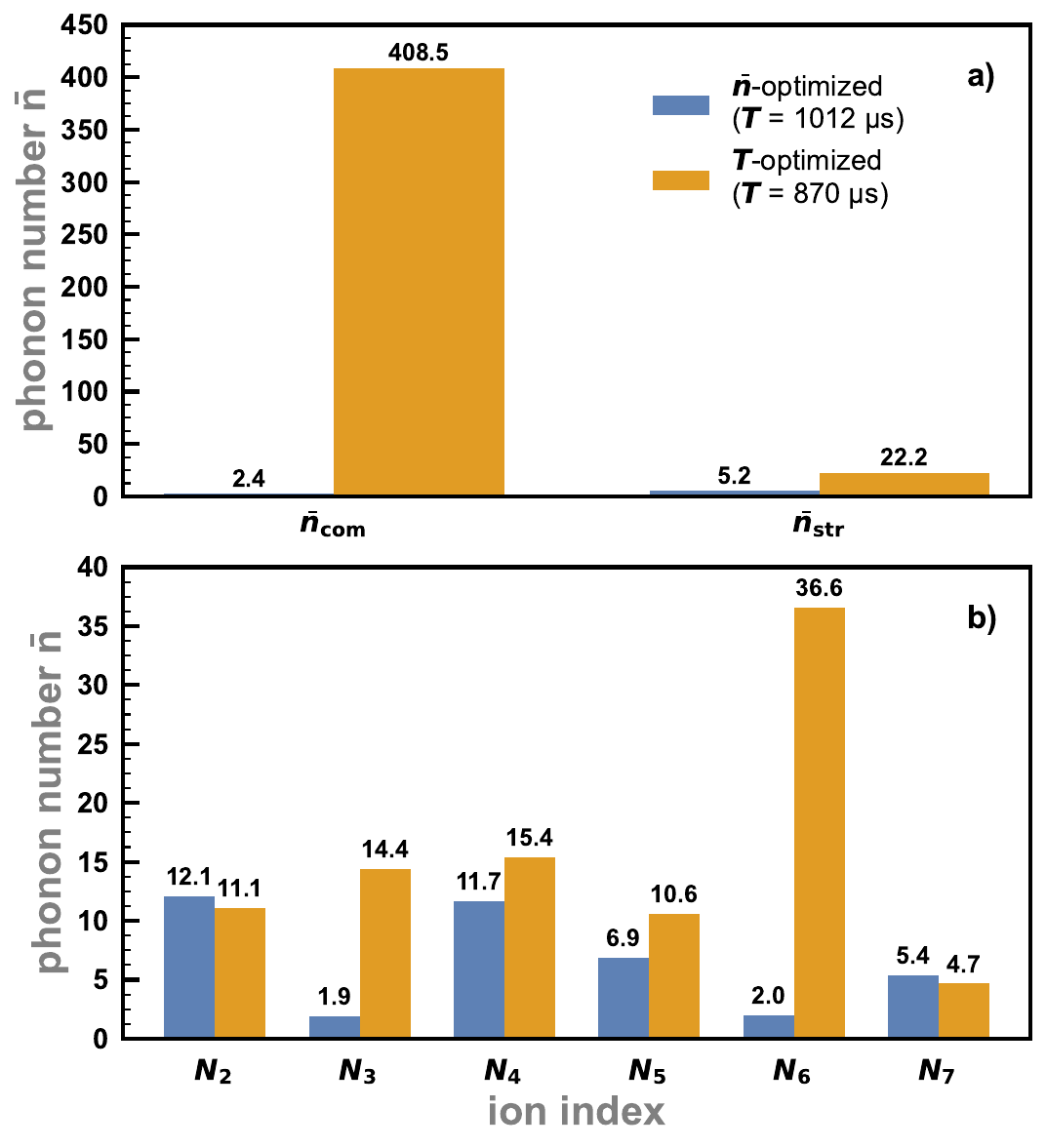}
    \caption{Phonon excitations for reorderings that map the ion configuration from Fig.~\ref{fig: layout}~\textit{a}) to an arbitrary configuration in which the ions number 1 and 8 are merged into a two-ion crystal at the central position (the central position is shown in Fig.~\ref{fig: layout}~\textit{d})). The execution time of the primitives is set to result in $\bar{n}\approx1$. The excitations are compared between a reordering routine optimized for the average phonon number (\textit{blue}) and one optimized for execution time (\textit{orange}). \textit{a}) Resulting average phonon number for the target ions $N_1$ and $N_8$ (displayed as the crystal’s energy in the common and stretch modes). \textit{b}) Average phonon number of the remaining ions.}
    \label{fig:Result_1}
\end{figure}

Fig.~\ref{fig:Result_1}~a) shows the results for weighting with the execution time or the motional excitation of the target ions $N_1$ and $N_8$. The difference in phonon numbers demonstrates the effectiveness of the minimizer. Furthermore, since the minimization was only applied to the target ions, the motional excitation on the rest of the ions may be larger (see Fig.~\ref{fig:Result_1}~b)). While the effect of the minimizer appears large, the resulting total phonon number is very sensitive to the execution time of the primitives. In other words, variations in the execution times and the non-repeatability of the primitives can result in a different outcome. Repeating the reordering operation but adjusting the execution time of the primitives to result in $\bar{n}\approx0.1$ (here, the \textit{near-adiabatic} regime) led to a motional excitation of $\bar{n}<2$ for all ions and minimization approaches. In this case, we obtained a reduction of $\approx 90$ percent for both the $\bar{n}_{\mathrm{\ com}}$ and $\bar{n}_{\mathrm{\ str}}$, and a execution time difference of $\approx 98\mu s$. While the simulation in the \textit{near-adiabatic} regime highlights the importance of minimizing the individual transport-induced excitations of the primitives, the simulation in the \textit{noisy} regime emphasizes the significance of quantifying noise contributions at the compiler level. Although the transport execution time increases, achieving a solution with a lower phonon number may be decisive if it falls below the cooling threshold, as cooling could require significantly more time.

\begin{figure*}
    \centering
    \includegraphics[width=\textwidth]{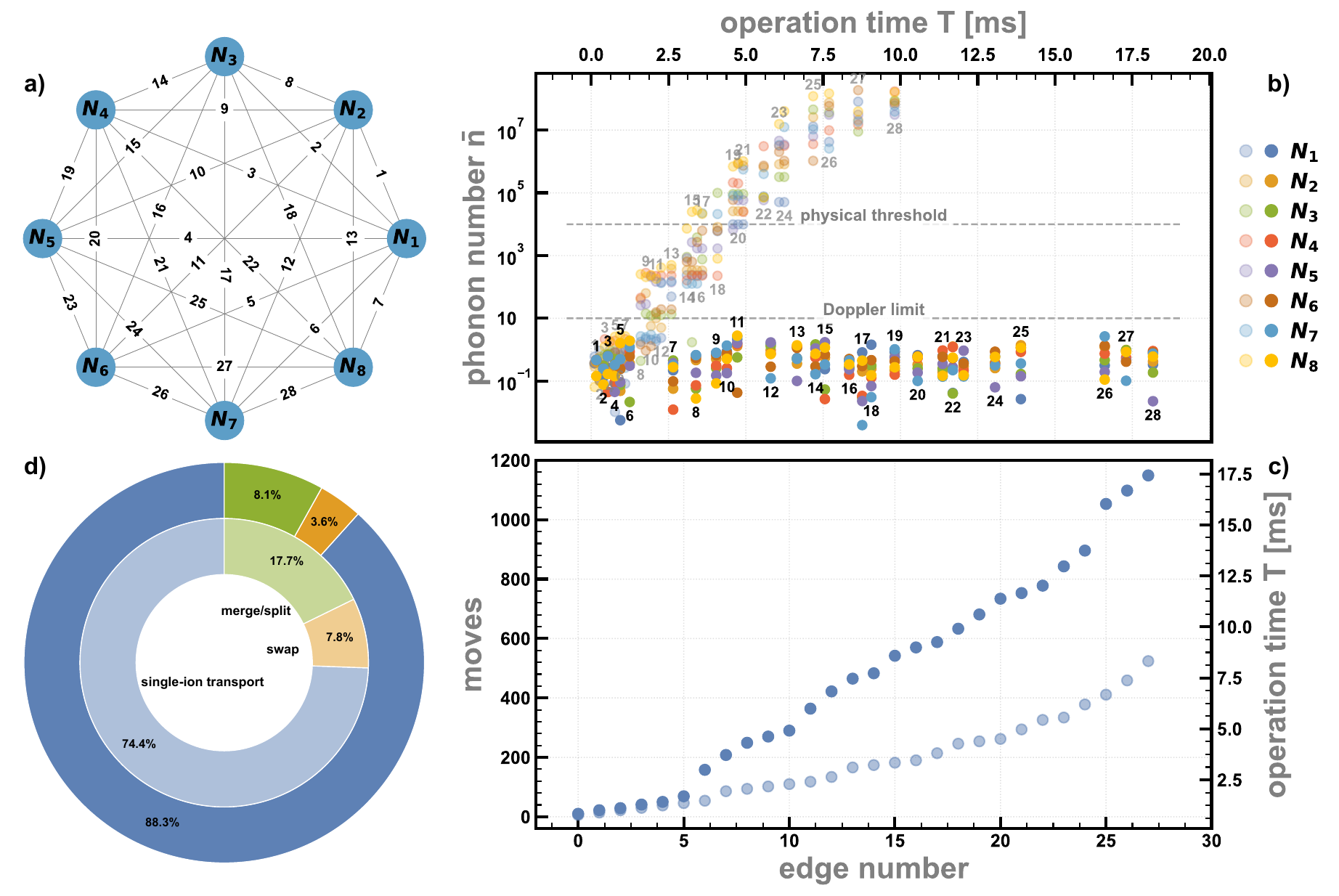}
    \caption{Simulation of ion reorderings (i.e., sequences of transport primitives) that map the ion configuration from Fig.~\ref{fig: layout}~\textit{a}) to different final configurations. Each of these final configurations is characterized by one specific pair of ions being merged into a two-ion crystal at the center of the gate zone. This prepares each pair for a two-qubit gate at the central position, shown in Fig.~\ref{fig: layout}~\textit{d}). Transparent graphics always show the data resulting from approach \textbf{(A)}, minimizing the operation time and the solid graphics always show the results from approach \textbf{(B)}, minimizing the phonon excitation. \textit{a}) Each possible ion pair is labeled with an edge number. \textit{b}) Accumulated phonon excitation for sequentially applied ion reorderings. The top $x$-axis shows the total operation time $T$. The edge numbers are directly assign to the data sets, indicating the last ion pair for which a two-qubit gate preparation was arranged. The individual primitive operations are all adjusted to $\bar{n}_s \approx 0.1$ according to Table~\ref{table:values_compiler}. \textit{c}) Accumulated number of \textit{moves} (a \textit{move} represents an arbitrary instruction from Table~\ref{tab:instructions}) and time as a function of the edge number corresponding to \textit{b}). \textit{d}) Distribution of the operation time across the individual primitive operations.}
    \label{fig:all-to-all}
\end{figure*}

However, it is important to mention that these results change with the initial configuration of the ions. This becomes important when considering running a given quantum circuit. Starting with the same configuration, we next estimate the required time and noise produced to implement two-qubit gates between arbitrary qubits\footnote{Respectively, simulate all-to-all connectivity.}. Meaning, we sequentially bring different pairs of ions to the gate zone, where they are merged into a two ion-crystal. Figure \ref{fig:all-to-all}~a) introduces a label for each set of operations required to pair a given couple of ions in the interaction zone (see caption). The light blue circles indicate the ion number, and the lines between them mark the corresponding set of operations used to pair these two ions in the gate zone. I.e., each line connecting two ions carries a label (the \textit{edge number}) which is subsequently used to identify that particular set of operations. For example, the line (set of operations) with edge number ``23'', brings the ions number 6 and 5 to the center of the gate zone and merges them into a crystal. The edge numbers are set in order of increasing operation complexity (number of moving steps) if one were to start from the configuration in Fig.~\ref{fig: layout}. In Fig.~\ref{fig:all-to-all}~b) we observe the collective evolution in time\footnote{After the $i$-th edge number is applied, the $(i+1)$-th set of operations is initialized using the outgoing ion configuration, operation time, and individual phonon numbers.} and average phonon number, based on the transport sequences calculated by the compiler. The uneven spacing between samples indicates the difference in execution time for different edge numbers. We present the results for two different searching approaches. First, \textbf{(A)} a search that minimizes the execution time with a large global boundary ($\approx10^8$) on the phonon number. Second, \textbf{(B)} a search that minimizes the maximum phonon number on every ion. The results obtained by optimizing with a global boundary \textbf{(A)} (\textit{transparent markers}) show a progressive increase in phonon number until the boundary is reached. However, after the pairing of ion 4 and 3 (edge ``14'') the estimated phonon number is outside the limits of our physical threshold (\textit{dashed line}). Therefore, after the physical threshold, these results do not represent a real-world quantity and are only qualitative. We determine this threshold from the minimum admissible value of $d(t)$, in a two-ion crystal with  $\nu_{\rm{COM}}=\SI{1}{\text{MHz}}$. This limit in phonon number is particularly understandable in a two-ion swapping operation, where the elongation of each ion approaches half the desired distance between the equilibrium positions\footnote{The equilibrium distance in a purely harmonic potential is $d =
\left(
\frac{q^2}
{2\pi \epsilon_0 m \omega_{\mathrm{COM}}^2}
\right)^{1/3}$
with $\omega_\mathrm{COM}=2\pi \times\SI{1}{\text{MHz}}$. The ions in these equilibrium positions, $\pm d/2$, cannot have an elongation, $A_{\text{tot}}$, larger than d/2.  The maximum phonon number can be found using Eq.~\eqref{5.12} with $x_{\mathrm{tot}}=x_{\mathrm{hom}}+x_{\mathrm{inh}}=A_{tot} \cos(\omega_\mathrm{COM}t+ \vartheta_{tot})$, resulting in  $\bar{n}\approx10^4$.}, contradicting our assumptions and introducing larger anharmonicities. The phonon excitation corresponding to this maximum ion elongation during the oscillation is approximately $10^4$ phonons. The trap depth for a single ion sets a final upper bound at $10^7$ phonons.\\
Following the second approach \textbf{(B)} (\textit{solid markers}), it is not only possible to stay within this limit but also to remain below the Doppler limit. This becomes possible by making use of destructive interference between the phases of the different ions. The solver uses information about the phases to find paths where destructive interference occurs. The magnitude of  interference depends on the $i-1$-th phase of each ion and the $i$-th operation. The variating magnitude is observed in Fig.~\ref{fig:all-to-all} as small oscillations in the phonon  number after completing a reordering step. This highlights the advantage of considering the phonon number for larger compiling processes. Overall, these results show that regardless the number of operations, it is possible to selectively minimize either the operation time or the phonon excitation of the targeted gate ions. However, one needs to be aware that timing errors in the experimental implementation can lead to very different results.

\subsection{Anomalous heating} \label{sec:anomalous}
The electrical noise present on an electrode's surface induces excitations in the modes of motion~\cite{deslauriersScalingSuppressionAnomalous2006,blakestad_transport_2010}. These effects become larger in surface-electrode traps as the ion-to-electrode distance decreases. This noise is known as anomalous heating with a rate given by: $$\dot{\bar{n}} = S_E(\omega) \, \frac{e^2}{4 m \hbar \omega} \quad .$$ The spectral noise density, $S_E(\omega)$, depends on the electrode geometry, the target trapping frequency, the trap temperature~\cite{deslauriersScalingSuppressionAnomalous2006,labaziewiczSuppressionHeatingRates2008}, and the passive electrical components attached to the electrode configuration~\cite{deslauriersScalingSuppressionAnomalous2006,kaushalShuttlingbasedTrappedionQuantum2020a}. Hence, heating rates are specific for each trap configuration and experimental setup. To have more realistic predictions, contributions from anomalous heating need to be added to the above discussed model. The total mean motional quantum number can be estimated as~\cite{kaufmannScalableQuantumProcessor2017}: $$\bar{n}_{\mathrm{tot}}\approx \bar{n}_{s}+\int \dot{\bar{n}} \, dt \quad .$$ While for short transport times motional excitation in the axial modes is dominated by the distribution of accelerations, anomalous heating represents the larger contribution for longer time scales. In addition, since, except for swapping operations, the direction of transport is typically perpendicular to the radial modes, anomalous heating remains the predominant source of motional excitation in the radial modes. As there are no generic models for anomalous heating, these need to be experimentally characterized. Using measurements for different radial and axial trapping frequencies allow us to create a specific heating model for our trap. This second model complements the discussed transport-induced excitations. There are other sources of electrical noise that can contribute to the existing heating. For example, a real DAC may introduce noise on the trap electrodes at a specific frequency range. However, the induced heating can be minimized by operating at a certain frequency regime, off resonant to the trapped ion. This could be controlled for example by carefully selecting the transport speed.

\section{Outlook} \label{sec:summary}
The methods developed here can be useful to optimize the performance of trapped-ion quantum processors by considering the impact that shuttling operations have on motional state occupation. For an actual algorithm, the compiler would need to optimize the sequence not only for a specific pair of target qubits in the gate zone, but for entire sequences involving two-qubit gates on all qubits and potentially including thresholds for sympathetic recooling. The specific implementation in a given quantum processor needs to account for other sources of noise, such as anomalous motional heating (whether technical or surface-related). Moreover, this model considers the pseudopotential approximation to calculate $\bar{n}$. To obtain a more realistic picture, we could instead consider the RF potential as laid out in Appendix~\ref{app:fundamental_matrix}. Using the RF potential might be necessary when modelling complex 2D structures that require through-junction transport or swapping at the junction centre. Ultimately, the compiler will need to assess the impact on performance for the entire algorithm. This may depend on the specific choice of qubit and gate mechanism. For example, the use of a long-lived first order magnetic-field insensitive qubit~\cite{langerLongLivedQubitMemory2005} may favor sequences that minimize motional heating, whereas lower internal-state coherence times may favor overall speed. The example chip design discussed here is intended for microwave near-field two-qubit gates~\cite{ospelkausTrappedIonQuantumLogic2008a,ospelkausMicrowaveQuantumLogic2011c}. These gates employ radial modes of motion for the two-qubit gates which are not affected by transport operation to the same extent that axial modes are and are also less affected by anomalous heating because of the generally higher frequency. The main effect on gate fidelity could be from cross-Kerr nonlinearities which modify the effective radial frequency for the gate mode as a function of the occupation number of the axial mode of motion~\cite{homeNormalModesTrapped2011}.

\section*{Acknowledgments} \label{sec:acknowledgements}
We acknowledge funding from the Ministry of Science and Culture of Lower Saxony (MWK) and the Volkswagen Foundation through the Quantum Valley Lower Saxony (QVLS) Q1 project, from the Federal Ministry for Research, Technology and Space (BMFTR) through the MIQRO and ATIQ projecs as well as the clusters4future project QVLS-iLabs SiQT, from the European Union through the Millenion-SGA1 project (101114305), and from DFG through the Cluster of Excellence QuantumFrontiers (EXC 2123).

\bibliography{qc}

\appendix

\section{Transport induced phonon excitation via propagator and generating functions}\label{sec:appendix_propagator}
Following \cite{husimiMiscellaneaElementaryQuantum1953, huculTransportAtomicIons2008}: Starting from the Schrödinger equation in Eq.~\eqref{schroedinger}, a further coordinate transformation, $u = s - \xi(t)$, together with the phase transformation
$\psi(u,t)=\phi(u,t)\exp\left(\frac{im}{\hbar}u\,\dot{\xi}\right)$, yields:
\begin{equation}
\begin{aligned}\label{appendixA1}
    i\hbar\frac{\partial\phi(u,t)}{\partial t}= &- \frac{\hbar^2}{2m}\frac{\partial^2\phi(u,t)}{\partial u^2}+\frac{1}{2}m\omega^2(t)u^2\phi(u,t)\\
    &+ m\left(\ddot{\xi}+\omega^2(t)\xi-\frac{f(t)}{m}\right)u\phi(u,t)\\
    &-m/2\left(\dot{\xi}^2+\dot{x}_0^2-\omega^2(t)\xi^2+\frac{2f(t)\xi}{m}\right)\phi(u,t)\\
\end{aligned}  
\end{equation}
The first line of Eq.~\eqref{appendixA1} can be identified as the undriven quantum mechanical oscillator with solution $\zeta(u,t)$. If we choose $\xi$ to be the classical solution of the driven harmonic oscillator (i.e., to obey $\ddot{\xi}=\frac{f(t)}{m}-\omega^2(t)\xi$), the second line in Eq.~\eqref{appendixA1} is satisfied. The final line in Eq.~\eqref{appendixA1}, which is independent of $u$, can be accounted for by another phase transformation, leading to:
\begin{equation}\label{appendixA2}
\psi(u,t)=\zeta(u,t)\exp\left[\frac{im}{\hbar}\dot{\xi}u+\frac{i}{\hbar}\int_{t_0}^t\mathcal{L}+\frac{m}{2}\dot{x}_0^2 \ dt\right]\,,
\end{equation}
with $\mathcal{L}$ the classical Lagrangian, $\mathcal{L}=\frac{m}{2}\dot{\xi}^2-\frac{m}{2}\omega(t)^2\xi^2+f\xi$, and $f=-m\ddot{x}_0$. Thus, the problem of the quantum mechanical forced parametric oscillator can be reduced to that of a quantum mechanical parametric oscillator and a classical forced parametric oscillator.
\\
The Schrödinger equation only provides a specific solution to the problem. However, to analyze the heating of the transported ion, we need to describe the general time-dependent dynamics of the system for arbitrary initial states in terms of the propagator $K(x,t|x',t')$, which solves the Schrödinger equation in operator form and provides the time evolution for any arbitrary state:
\begin{align}\label{appendixA3}
    \psi(x,t)=\int K(x,t|x',t')\psi(x',t')\ \, dx'\, .
\end{align}
This then provides the desired transition amplitudes $a_{nk}$ between the eigenstates $\psi_k$ and $\psi_n$ of the Hamiltonian:
\begin{align}\label{appendixA4}
    a_{nk}(t,t')=\iint\psi^*_n(x)K(x,t|x',t')\psi_k(x')\ \, dx \, dx'
\end{align}
The transition probabilities are then given by $P_{nk}(t,t') = |a_{nk}|^2$. The propagator can be calculated using the Feynman path integral formalism and generalized for the harmonic oscillator:
\begin{equation}
\begin{aligned}\label{appendixA5}
    &K_{\text{HO}}(x_b,t_b|x_a,t_a)=\sqrt{\frac{m\omega_0}{2\pi i\hbar\sin{(\omega_0\Delta t)}}}\\
    &\hfill\cdot \exp\left[\frac{im\omega_0}{2\hbar\sin{(\omega_0\Delta t)}}((x_a^2+x_b^2)\cos{(\omega_0\Delta t)}-2x_bx_a)\right]\,
\end{aligned}
\end{equation}
with $\Delta t=(t_b - t_a)$. The propagator in the parametric case is then obtained by replacing the trigonometric expressions in \eqref{appendixA5} (i.e., the solutions of the free harmonic oscillator $\ddot{x}+\omega_0^2 x=0$) with $X$ and $Y$, the solutions of the classical parametric oscillator. This yields the corresponding propagator:
\begin{equation}
\begin{aligned}\label{appendixA6}
    &K_{\text{PHO}}(x_b,t_b|x_a,t_a)=\sqrt{\frac{m}{2\pi i\hbar X(t_b,t_a)}}\\
    &\cdot \exp\left[\frac{im}{2\hbar X(t_b,t_a)}(x_b^2\dot{X}(t_b,t_a)-2x_bx_a+x_a^2Y(t_b,t_a))\right]
\end{aligned}
\end{equation}
(It is shown in \cite{husimiMiscellaneaElementaryQuantum1953} that this is indeed the case.) To perform the final step and obtain the propagator in the driven parametric case, one can use the structural relationship derived for the corresponding Schrödinger equation in Eq.~\eqref{appendixA2}:
\begin{equation}
\begin{aligned}\label{appendixA7}
    K_{\text{FPHO}}(x_b,t_b|x_a,t_a)&=K_{\text{PHO}}(x_b,t_b|x_a,t_a)\\&\cdot\exp\left[\frac{im}{\hbar}\dot{\xi}x_b+\frac{i}{\hbar}\int_{t_a}^{t_b}\mathcal{L}+\frac{m}{2}\dot{x}_0^2 \ dt\right]
\end{aligned}
\end{equation}

\subsection{Excitation with an initial ground state preparation}
Lastly, one needs to find a closed form to evaluate the phonon excitation $\langle n\rangle=\sum_n nP_{n0}(t,t')$ (in this case assuming an initial preparation in the ground state). Therefore, one may use a power series approach, where the initial state $k$ and the final state $n$ can be selected based on the powers of $w$ and $v$:
\begin{align} \label{4.24}
P(v,w) = \sum_{k,n} v^k w^n P_{n,k}(t,t') = \sum_{k,n} v^k w^n |a_{n,k}|^2\, .
\end{align}
Substituting Eq.~\eqref{appendixA4} for the transition amplitudes into \eqref{4.24}, with $|a_{n,k}|^2 = a^*_{n,k} \cdot a_{n,k}$, yields:
\begin{equation}
\begin{aligned}\label{appendixA8}
    &P(v,w)=\iiiint K^*(x,t|x',t')K(y,t|y',t')\\
    &\cdot \sum_n w^n\psi^*_n(x)\psi_n(y) \sum_k v^k\psi^*_k(x')\psi_k(y')\ \, dx \, dx' \, dy \, dy'\, .
\end{aligned}
\end{equation}
This can be simplified by expressing the eigenfunctions as:
\begin{equation}
\begin{aligned}\label{appendixA9}
    \psi_n(y)\psi_n(x)&=\left(\frac{m\omega_0}{\pi\hbar}\right)^{\frac{1}{2}}\frac{1}{2^nn!}H_n(\alpha y)H_n(\alpha x)\\
    &\exp\left(-\frac{\alpha^2}{2}(y^2+x^2)\right)\,,
\end{aligned}
\end{equation}
with $\alpha = \sqrt{\frac{m\omega_0}{\hbar}}$. Using the Mehler kernel for the Hermite polynomials $H_n$:
\begin{equation}
\begin{aligned}\label{appendixA10}
    \sum_{n=0}^{\infty}\frac{v^n}{2^nn!}&H_n(\alpha y)H_n(\alpha x)=\\
    &\frac{1}{\sqrt{1-v^2}}\exp\left(\alpha^2\frac{2vxy-v^2(x^2+y^2)}{1-v^2}\right)\,,
\end{aligned}
\end{equation}
and multiplying with the factor $\frac{\alpha}{\sqrt{\pi}}\exp\left(-\frac{\alpha^2}{2}(y^2+x^2)\right)$ from both sides then yields:
\begin{equation}
\begin{aligned}\label{appendixA11}
    \sum_{n=0}^{\infty}&v^n\psi_n(y)\psi^*_n(x)=\\
    &\frac{\alpha}{\sqrt{\pi(1-v^2)}}\exp\left(-\alpha^2\frac{(1+v^2)(y^2+x^2)-4vyx}{2(1-v^2)}\right)
\end{aligned}
\end{equation}
Substituting \eqref{appendixA11} in \eqref{appendixA8} and applying the Gaussian integral formula gives:
\begin{equation}
\label{appendix12}
\begin{aligned}
P(v,w)
&= \sqrt{\frac{2}{D(t,t_0)}} \\
&\quad\times
\exp\left[
-\frac{(1-v^2)(1-w^2)}{D(t,t_0)}
\left(
\eta(t,t_0)\frac{1-v}{1+v}
\right.\right.\\
&\hspace{5.5em}\left.\left.
+\,\eta(t,t_0)\frac{1-w}{1+w}
\right)
\right]\\
&= \sum_{k,n}v^k w^n P_{n,k}(t,t_0)\,,
\end{aligned}
\end{equation}
with
\begin{equation}
\begin{aligned}
D(t,t_0)&=Q(t,t_0)(1-v^2)(1-w^2)\\
&+(1+v^2)(1+w^2)-4vw\,,
\end{aligned}
\end{equation}
which leads to the final result:
\begin{equation}
\begin{aligned}\label{appendixA12}
    \langle n\rangle_0&=\frac{\partial P(0,w)}{\partial w}\Bigg|_{w=1}=\eta+\frac{1}{2}(Q-1)\,,
\end{aligned}
\end{equation}
with $Q(t,t_0)=\frac{1}{2}\left(\omega_0^2X^2+\dot{X}^2+Y^2+\frac{1}{\omega_0^2}\dot{Y}^2\right)$.

\subsection{Excitation starting from an arbitrary initial Fock state}
\label{sec:appendix_initial_fock}

Extracting $\langle n\rangle_0$ in \eqref{appendixA12} required differentiating $P(v,w)$ with respect to $w$ at $w=1$ and evaluating at $v=0$. If the initial excitation $k'$ differs from the ground state, $k'\neq0$, one must instead first differentiate $k'$ times with respect to $v$. This yields terms proportional to $v^{\,k-k'}$, so that setting $v=0$ afterwards leaves only the summand with $k=k'$:
\begin{equation}
\label{appendixA13}
\begin{alignedat}{2}
&\frac{1}{k'!}\frac{\partial^{k'}}{\partial v^{k'}}
\left[\frac{\partial P(v,w)}{\partial w}\right]\Bigg|_{v=0,\,w=1}
\\[2pt]
&\qquad = \frac{1}{k'!}\frac{\partial^{k'}}{\partial v^{k'}}
   \left[\sum_{k,n}v^k w^n P_{n,k}(t,t')\right]\Bigg|_{v=0,\,w=1}\\[2pt]
&\qquad = \frac{1}{k'!}\frac{\partial^{k'}}{\partial v^{k'}}
   \left[\sum_k v^k \langle n\rangle_k\right]\Bigg|_{v=0}\\[2pt]
&\qquad = \frac{1}{k'!}\sum_k
   \underbrace{k(k-1)\cdots\big(k-(k'-1)\big)}_{k'\ \text{factors}}
   v^{\,k-k'}\,\langle n\rangle_k \Bigg|_{v=0}\\
&\qquad =\; \langle n\rangle_{k'}\,.
\end{alignedat}
\end{equation}
with $\langle n\rangle_k=\sum_n n\,P_{n,k}(t,t')$. This is the expectation value
of the phonon excitation for the transitions from the initial state $k'$ to all
other accessible states, induced either by the acceleration of the potential or
by a change in the harmonic trap frequency. For $P(v,w)$ from
\eqref{appendix12}, the power series representation reads
\begin{multline}
\label{appendixA14}
\frac{\partial P}{\partial w}\Bigg|_{w=1}
= \sum_{k\geq0}
\left(\eta-\frac{1}{2}\left[
(1-Q)\binom{k+2}{2}\right.\right. \\
\left.\left.
-2\binom{k+1}{2}
+(1+Q)\binom{k}{2}
\right]\right)v^k .
\end{multline}
whose coefficients simplify to $\eta+\frac{1}{2}(Q-1)+kQ$, so that
\begin{equation}
\label{appendixA15}
\langle n\rangle_{k'}\;=\;\langle n\rangle_0+k'Q\, .
\end{equation}

The contribution of the acceleration is thus still provided by the semiclassical analogy via $\eta$ and is independent of the initial state, whereas the impact of the frequency change entails an additional term scaling with the initial phonon number $k'$.

\section{Parametric excitation by time-dependent mode frequencies}
\label{sec:appendix_bogoliubov}

A time-dependent trap frequency can excite the ion even when it is never displaced. This parametric contribution follows directly from the second moments of the initial ground state, $\langle x^2\rangle=\hbar/2m\omega_i$, $\langle\dot{x}^2\rangle=\hbar\omega_i/2m$ and $\langle x\dot{x}+\dot{x}x\rangle=0$. Since the dynamics is linear, these propagate according to the fundamental solutions $X$ and $Y$ known from Appendix~\ref{sec:appendix_propagator}, $x(T)=Y x_0+X\dot{x}_0$. Inserting
them into $E=\frac{m}{2}\langle\dot{x}^2\rangle
+\frac{m}{2}\omega_f^2\langle x^2\rangle$ gives, in units of $\hbar\omega_f$,
\begin{equation}
\label{appendixB9}
    Q=\frac{1}{2}\left[\omega_i\omega_f X^{2}
      +\frac{\omega_i}{\omega_f}\dot{X}^{2}
      +\frac{\omega_f}{\omega_i}Y^{2}
      +\frac{\dot{Y}^{2}}{\omega_i\omega_f}\right],
\end{equation}
with $\omega_i=\omega(0)$ and $\omega_f=\omega(T)$. $Q$ is thus the final energy measured in units of the zero-point energy of the final trap, and the excitation out of the ground state is $\frac{1}{2}(Q-1)$. For a constant frequency, the fundamental solutions are $X=\sin(\omega_0T)/\omega_0$ and $Y=\cos(\omega_0T)$, giving $Q=\sin^2(\omega_0T)+\cos^2(\omega_0T)=1$. In the adiabatic limit one finds $Q\rightarrow1$ as well, irrespective of the net change between $\omega_i$ and $\omega_f$.

\begin{figure}
    \centering
    \includegraphics[width=\columnwidth]{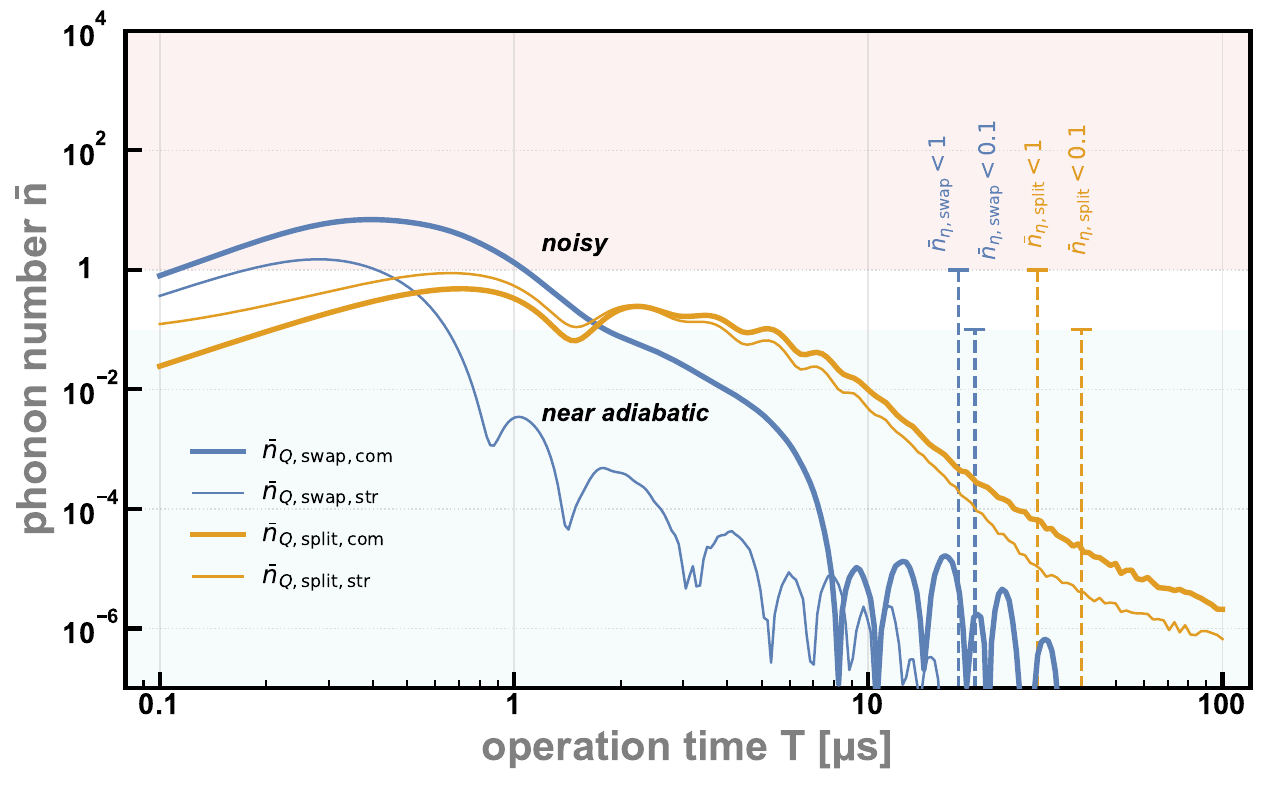}
    \caption{Mean parametric phonon excitation out of the motional ground state, caused by the time dependence of the mode frequencies $\bar{n}_Q$. Shown are the excitations of the primitive operations “two-ion swapping” and “two-ion splitting”, where “two-ion splitting” is representative of “two-ion merging”, as the graphs were almost identical in this plot. A comparison with the acceleration-induced phonon excitations at the times of interest ($1,\mu$s and $0.1,\mu$s) for the compiler operation (see Table~\ref{table:values_compiler}) is also given.}
    \label{fig:appendixB}
\end{figure}

Fig.~\ref{fig:appendixB} shows the parametric phonon excitation out of the motional ground state, caused by the time dependence of the mode frequencies $\bar{n}_Q=\frac{1}{2}(Q-1)$. At the times extracted in Table~\ref{table:values_compiler} for the compiler evaluation, $\bar{n}_Q$ is remarkably small compared to the acceleration-induced excitation $\eta$ and is therefore safely neglected throughout this work.

\section{The anharmonic correction to the quantum--classical correspondence}
\label{appendixC}

The quantum--classical correspondence assumed throughout this work is valid for harmonic potentials. In the presence of anharmonic potentials the classical description deviates from the full quantum mechanical model. We integrated the latter using \textsc{QuTiP}, in a fluctuation frame. Since the classical motion in the trap frame passes through high motional states during the transport, even if converging to lower $\bar{n}$ at the end of the operation ($t=T$), representing such a state directly would require a Fock space of comparable dimension. Moving with the classical trajectory instead separates the quantum correction from the classical background and leaves a state that stays close to the ground state, so that a truncation at $60$ Fock states proved to be sufficient. Furthermore, the calculation is carried out using natural units to avoid numerical errors.

With the time-dependent unitary
\begin{equation}
\label{appendixC1}
    \hat U(t)=\exp\!\Big[\tfrac{i}{\hbar}\big(m\dot\xi\,\hat x-\xi\,\hat p\big)\Big] ,
\end{equation}
where $\xi$ is the classical trajectory in the frame of the moving trap, in which the transport enters as the inertial force $f=-m\ddot x_0$ of Appendix~\ref{sec:appendix_propagator}, one has $\hat U^{\dagger}\hat x\hat U=\hat x+\xi$ and $\hat U^{\dagger}\hat p\hat U=\hat p+m\dot\xi$. Therefore $\hat U$ is the displacement operator $\hat D$ with amplitude $\sqrt{m\omega/2\hbar}\,(\xi+i\dot\xi/\omega)$ when referred to a mode of frequency $\omega$. Writing $\hat x=\xi(t)+\hat u$ and expanding the potential up to quartic order, every term linear in $\hat u$---including the term $m\ddot\xi\,\hat u$ generated by $\hat U$ and the inertial force $-f\hat u$---collects into
\begin{equation}
\label{appendixC2}
    m\,\hat u\,\Big(\underbrace{\ddot\xi+\omega^{2}\xi
      +a_h\xi^{2}+b_h\xi^{3}-\tfrac{f}{m}}_{=\,0}\Big)=0\ ,
\end{equation}
which vanishes when $\xi$ obeys the classical equation of motion \emph{including} the anharmonic forces. The classical drive is thus removed exactly, and what remains is the Hamiltonian of $\phi$, the state in the co-moving frame introduced in \eqref{appendixA1},
\begin{multline}
\label{appendixC3}
    \hat H_\phi=\frac{\hat p^{2}}{2m}
    +\tfrac12 m\omega^{2}\hat u^{2}
    +\underbrace{m\big(a_h\xi+\tfrac32 b_h\xi^{2}\big)\hat u^{2}}_{\text{shifts }\omega_{\rm eff}} \\
    +\underbrace{m\big(\tfrac13 a_h+b_h\xi\big)\hat u^{3}
    +\tfrac14 m\,b_h\,\hat u^{4}}_{\text{genuinely non-Gaussian}} .
\end{multline}
Only the anharmonicity couples to the classical trajectory here: in a purely harmonic trap both braces vanish and the quantum--classical correspondence is restored.

Among our transport primitives, merging and splitting exhibit comparatively large anharmonicities in the potential. Therefore the case of the splitting is used here to apply the method\footnote{Therefore $d(t)$ and $\mu$ will replace $x_0$ and $m$ in \eqref{appendixC1}--\eqref{appendixC3}.}. Expanding the trap potential $q(\alpha x^2+\beta x^4)$ together with the Coulomb term about the instantaneous equilibrium separation $d(t)$ gives, in terms of the relative coordinate $x_1-x_2$\footnote{Of the two axial modes, only the stretch mode is considered, so $\omega$ in Eq.~\eqref{appendixC3} denotes $\omega_{\rm str}$ of Eq.~\eqref{eq: STRvsCOM}. The centre-of-mass mode is omitted since the trap remains symmetric about the origin.} with reduced mass $\mu=m/2$,
\begin{equation}
\label{appendixC4}
    a_h=\frac{3q\beta d}{2\mu}-\frac{3q^{2}}{4\pi\epsilon_0\mu\,d^{4}},
    \qquad
    b_h=\frac{q\beta}{2\mu}+\frac{q^{2}}{\pi\epsilon_0\mu\,d^{5}} ,
\end{equation}
where $\alpha(t)$ and $\beta(t)$ follow from $\omega_{\rm com}$ together with the equilibrium condition \eqref{eq:position merging}, $\beta=\big[m\omega_{\rm com}^{2}/q-2q/(4\pi\epsilon_0 d^{3})\big]/(2d^{2})$.

The two braces in \eqref{appendixC3} are separated by propagating the Hamiltonian three times: with the anharmonic terms switched off entirely, with the first brace alone, and complete. The first case reduces to the parametric problem of Appendix~\ref{sec:appendix_bogoliubov} and reproduces the results of \eqref{appendixB9}, which validates the scheme. The second yields $\delta\bar n_Q$, the shift of the parametric excitation caused by the anharmonic correction to $\omega_{\rm eff}$; since $\bar n_Q$ is discarded in Appendix~\ref{sec:appendix_bogoliubov} in the first place, so is its shift. What remains, $\Delta\bar n_q^{\rm anh}$, is the genuinely non-Gaussian contribution of the cubic and quartic terms. Because the eigenstates of an anharmonic potential are no longer Fock states, the excitation is evaluated as $\bar n=\sum_k k\,|\langle\chi_k(T)|\phi(T)\rangle|^2$ in the eigenbasis $\chi_k$ of the final Hamiltonian, which reduces to the phonon number in the harmonic limit\footnote{The excitation is referred to the ground state of the \emph{final} potential. This matters because $\omega_{\rm str}$ changes during the operation: taking the initial potential as the reference would add a static offset to $\bar n$, which reflects the frequency change alone and not the transport.}.

\begin{figure}
    \centering
    \includegraphics[width=\linewidth]{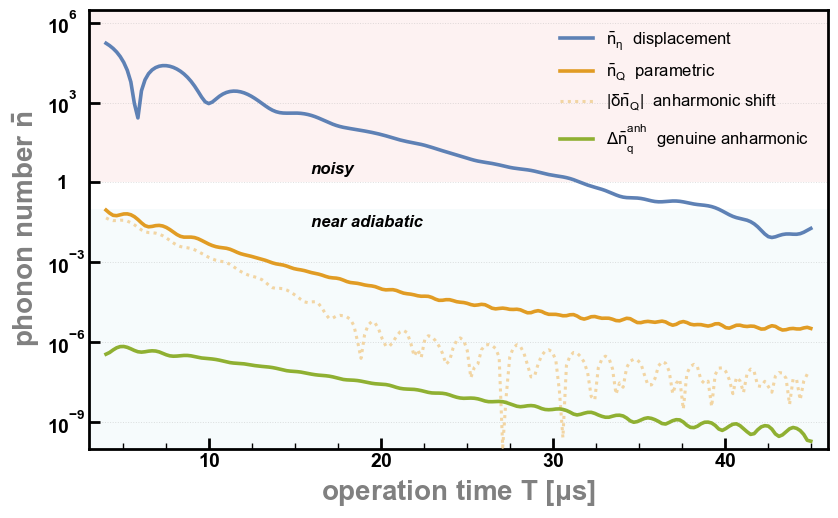}
    \caption{Anharmonic quantum correction of the splitting primitive versus the operation duration $T$. The two anharmonic contributions $|\delta\bar n_Q|$ and $\Delta\bar n_q^{\rm anh}$ generated by the two braces of \eqref{appendixC3} are shown alongside the displacement contribution $\eta$ and the parametric contribution $\bar n_Q$ of Appendix~\ref{sec:appendix_bogoliubov}.}
    \label{fig:anharmonic_mergesplit}
\end{figure}

The results can be seen in Fig.~\ref{fig:anharmonic_mergesplit}. The genuine anharmonic contribution $\Delta\bar n_q^{\rm anh}$ is orders of magnitude smaller than the displacement contribution $\eta$ in the case of two-ion splitting. Nor could any relevance of the quantum correction in the case of the other primitives be found.

\section{Linearisation with the full radio-frequency field}
\label{app:fundamental_matrix}

Eq.~\eqref{eq:rot equations of motion pot} considers the pseudopotential in the definition of $\Phi_{\rm{tot}}$. However, in this appendix, we use the time-dependent total potential. This can be expressed by expanding the equation $\ddot{\mathbf r}=\mathbf F(\mathbf r,t)$, where $\mathbf F(\mathbf r,t)$ is the force divided by the ion mass, along the reference trajectory $\mathbf r_{\mathrm{ref}}$:
\begin{equation}
  \ddot{\mathbf w}
  \;=\;
  \underbrace{
    \mathbf b(t)
  }_{\substack{
    \mathbf F(\mathbf r_{\mathrm{ref}},t)-\ddot{\mathbf r}_{\mathrm{ref}}
  }}
  \;-\;
  \underbrace{
    H(t)\,\mathbf w
  }_{\substack{
    -\left.\partial\mathbf F/\partial\mathbf r\right|_{\mathbf r_{\mathrm{ref}}}\mathbf w
  }}
  \;+\;
  \mathcal O\!\left(|\mathbf w|^{2}\right).
  \label{eq:app_expansion}
\end{equation}
Where $\mathbf r\in\mathbb R^{n}$ collects all laboratory-frame coordinates of the two ions and $\mathbf w=\mathbf r-\mathbf r_{\mathrm{ref}}$ is the deviation from the reference trajectory, carrying the motional state. The effects of the RF enter both terms: the RF force at the reference point, $\mathbf f_{\mathrm{rf}}(\mathbf r_{\mathrm{ref}})\cos(\Omega_{\mathrm{rf}}t)$, is present in $\mathbf b$, and the curvature of the RF potential is contained in $H$. The equation is still linear, with:
\begin{equation}
  H(t)\;=\;H_{\mathrm{dc}}\big(\theta(t)\big)
        \;+\;H_{\mathrm{rf}}\big(\theta(t)\big)\cos(\Omega_{\mathrm{rf}}t)
  \label{eq:app_Ht}
\end{equation}
Written in the phase-space state
$\boldsymbol\zeta=(\mathbf w,\dot{\mathbf w})^{\!\top}$, the homogeneous part of
\eqref{eq:app_expansion} therefore possesses a
fundamental matrix $\Theta(t,t_{0})$.

Equation~\eqref{eq:app_Ht} furthermore simplifies the evaluation of $\Theta$. The curvature $H$ does not have to be recomputed inside an integration resolving $\Omega_{\mathrm{rf}}$. This is because $H_{\mathrm{dc}}$ and $H_{\mathrm{rf}}$ depend on time only through the slow transport coordinate $\theta$, whereas the sole fast dependence, $\cos(\Omega_{\mathrm{rf}}t)$, is analytic. $H_{\mathrm{dc}}$ and $H_{\mathrm{rf}}$ are therefore sampled on a grid $\{\theta_{k}\}$ spanning the trajectory, and each of the $n\times n$ matrix elements is interpolated as a smooth function of $\theta$. Since the RF term is strictly proportional to $\cos(\Omega_{\mathrm{rf}}t)$, the two matrices follow from one evaluation each per grid point. During the time integration the curvature is then assembled according to Eq.~\eqref{eq:app_Ht}. The field evaluations are confined to the transport grid, while the integrator resolves $\Omega_{\mathrm{rf}}$ with steps that carry no field evaluation at all.\\

Since the solution map of our operations remains affine in the incoming state, the recursion of Eq.~\eqref{eq:recursive2} applies unchanged.

\end{document}